\documentclass[a4paper, 11pt]{article}
\usepackage[utf8]{inputenc}

\usepackage{bigints}

\usepackage[english]{babel}
\usepackage[T1]{fontenc}
\usepackage[babel=true]{csquotes} 
\usepackage{aeguill}

\usepackage{latexsym}
\usepackage{comment}
\usepackage{natbib}

\usepackage{tikz}
\usepackage{amsmath,amssymb}
\usetikzlibrary{shapes, arrows.meta, positioning, fit}

\usepackage{appendix}
\usepackage[normalem]{ulem}
\usepackage{upgreek}
\usepackage{tabularx}
\usepackage{amssymb}
\usepackage{amsmath}
\usepackage{amsmath,mathrsfs}
\usepackage{amsthm}
\usepackage{graphicx}
\usepackage{empheq}
\usepackage{color}
\usepackage{empheq}
\usepackage{float}
\usepackage{caption}
\usepackage{subcaption}
\usepackage[usenames,dvipsnames]{xcolor}
\usepackage{soul}
\usepackage{adjustbox}
\usepackage{listings}
\usepackage{algorithm}
\usepackage{algpseudocode}
\usepackage{bbm}
\usepackage{enumitem}
\usepackage{booktabs}
\usepackage{array}
\usepackage{multirow}
\newcolumntype{C}[1]{>{\centering\arraybackslash}p{#1}}
\usepackage{mathrsfs}

\definecolor{dkgreen}{rgb}{0,0.6,0}
\definecolor{gray}{rgb}{0.5,0.5,0.5}
\definecolor{mauve}{rgb}{0.58,0,0.82}

\def \R{\mathbb{R}}

\usepackage{amsmath, amssymb}
\usepackage{graphicx}
\usepackage{color}
\usepackage{textcomp}
\usepackage{graphics}
\usepackage{float}
\usepackage{empheq}
\usepackage{caption}
\usepackage{subcaption}
\usepackage{hyperref}
\usepackage{mathrsfs}

\usepackage{comment}

\usepackage{geometry}
\usepackage{stmaryrd}

\AtBeginDocument{}
\AtBeginDocument{}

\mathtoolsset{showonlyrefs}

\newcommand{\be}{\begin{align}}
	\newcommand{\ee}{\end{align}}

\newcommand{\bea}{\begin{eqnarray}}
	\newcommand{\bes}{\begin{subEquations}}
		\newcommand{\ees}{\end{subEquations}}
	\newcommand{\bgt}{\begin{gather}}
		\newcommand{\egt}{\begin{gather}}
			\newcommand{\eea}{\end{eqnarray}}
		\newcommand{\beaa}{\begin{eqnarray*}}
			\newcommand{\eeaa}{\end{eqnarray*}}

		\def\beqs{\begin{eqnarray*}}
			\def\enqs{\end{eqnarray*}}
		\def\beq{\begin{eqnarray}}
			\def\enq{\end{eqnarray}}

		\numberwithin{equation}{section}

		\graphicspath{{figures/}}

\begin{document}

			\title{The Fundamental Structure of Risk: From\\ Characteristics to Covariance
            }

			\author{
				Alexandre Alouadi \footnote{BNP Paribas CIB and Ecole Polytechnique, CMAP.  \sf alexandre.alouadi@polytechnique.edu}
				\and Charles-Albert Lehalle \footnote{Ecole Polytechnique, CMAP. \sf charles-albert.lehalle@polytechnique.edu}
			}

			\date{}
			\maketitle

\begin{abstract}

    Estimating the covariance structure of financial assets typically relies on historical returns,
    making risk models dependent on noisy and asset-specific time series. We propose the
    Characteristic-Driven Dynamic Factor Model (CD-DFM), a non-linear latent factor model that instead
    constructs a representation of the asset cross-section directly from observable firm characteristics,
    primarily company fundamentals.
    The learned latent space jointly determines interpretable factor exposures and a forward covariance
    estimator, and is trained end to end on an objective that combines a Stein covariance loss with a
    factor reconstruction term, targeting the out-of-sample second moments used in risk management.
    Because the latent representation, i.e.\ the encoder depends only on
    characteristics, previously unseen assets can be embedded at inference time without retraining.
    Experiments on S\&P 500 equities show that CD-DFM produces economically structured
    latent representations, interpretable factor portfolios, and competitive covariance forecasts despite
    relying on substantially lower-frequency information than return-based approaches. Among the
    benchmarked methods, it is the only model that simultaneously combines characteristic-driven
    representations, factor interpretability, competitive covariance calibration, and zero-shot onboarding
    of unseen assets. The code is available at \textcolor{blue}{\href{https://github.com/alexouadi/CD-DFM}
{https://github.com/alexouadi/CD-DFM}}.

\end{abstract}

			\vspace{5mm}

			\noindent {\bf Keywords}: latent factor models, covariance estimation, firm characteristics,
			representation learning, equity risk

\section{Introduction}

Learning latent representations has become a standard paradigm for high-dimensional structured data.
The idea is that such data can be compressed into a low-dimensional latent space that captures its
meaningful variation while discarding noise
(beginning with Principal Component Analysis \cite{benzecri1977analyse},
continuing with nonlinear extensions such as Self-Organizing Maps \cite{deboeck2013visual},
and more recently machine-learning representations \citep{bengio2013}),
because structured data tends to have low
intrinsic dimensionality \citep{fefferman2016}: the manifold of valid observations occupies a tiny
fraction of the ambient space, and the latent embedding provides a coordinate system on it. The same
principle extends to time series, where forecasting in a learned latent space is more stable than in the
raw observation space \citep{yang2026}. This raises a natural question for financial assets: can we build a
meaningful latent representation of the cross-section, and use it to estimate risk? The setting is harder
than the usual one, since the signal-to-noise ratio is low \cite{welch2008comprehensive},
the cross-section is heterogeneous \cite{harvey2016and}, and there
is no ground truth for what a correct representation should look like.

The standard framework for this is a factor model \cite{ross2013arbitrage}, in which returns $r_t \in \R^n$ decompose as
$r_{t} = \beta f_t + \varepsilon_{t}$: the \emph{factors} $f_t \in \R^K$
capture common sources of risk shared across assets, the \emph{loadings}
$\beta \in \R^{n \times K}$ measure each asset's exposure to them, and $\varepsilon_{t} \in \R^n$ is the idiosyncratic noise.
The dominant approach for estimating this structure from data is the Principal Component Analysis (PCA),
which identifies factors as the projection of the returns onto the leading eigenvectors of the return covariance matrix.
PCA is statistically principled, but its factors are linear combinations of all assets
with no economic label, making interpretation difficult.
Sparse PCA \citep{zou2006} addresses this partly by constraining loadings to be sparse,
so each factor loads on a small subset of assets. Interpretability improves, but the
factors remain purely statistical constructs from the returns alone, as sparsity selects assets, not economic themes,
and the resulting factors can still vary arbitrarily across time.

The industry, in contrast, relies on characteristic-based factors. These are typically built in two
stages: a one-to-one mapping of each chosen characteristic to a factor (no compression), followed by a
PCA that linearly reduces the resulting dimension \cite{WynneBrownCeria2014,MSCIBarraGEM3}.

A more principled route is to link factor loadings to observable firm characteristics.
\citet{kelly2019} propose Instrumented PCA (IPCA), in which loadings are constrained to be
linear functions of characteristics, so $\beta_{i,t} = x_{i,t}^\top \Gamma, \, i \leq n$ for a matrix
$\Gamma$ to be estimated.
This yields time-varying, characteristic-driven loadings and a direct connection between
economic variables and the latent factor structure.
Note, however, that the ``characteristics'' used in this literature are not all fundamentals: they
often include market-driven metrics such as the beta or the return volatility of the companies considered.
The limitation is linearity: the mapping from characteristics to loadings is a single matrix
multiplication, which may be too restrictive to capture the nonlinear interactions that
drive asset co-movement in practice. \citet{gu2019}, hereafter GKX, address this by replacing the linear
mapping with a neural network.
Their conditional autoencoder learns $\beta_{i,t} = g(x_{i,t}), \, i \leq n$ where $g$ is an MLP,
and estimates factors jointly with loadings via an autoencoder objective.
The result is a nonlinear, characteristic-driven factor model that achieves smaller
out-of-sample pricing errors than its linear predecessors.
However, the latent space is not analysed for economic interpretability, and  GKX is trained to minimise return reconstruction error (pricing errors), not for forward
covariance prediction, which is a distinct and harder objective.

In our view, an ideal latent representation of financial assets should satisfy three properties.
\textbf{(P1) Characteristic-driven}: the representation is a function of observable firm
characteristics, so that similar stocks are close in the latent space and a new, unseen
asset can be embedded from its characteristics alone, without requiring past returns.
\textbf{(P2) Interpretable}: each latent factor corresponds to a recognisable economic
theme, not an arbitrary statistical direction.
\textbf{(P3) Risk-predictive}: the representation supports accurate out-of-sample covariance
estimation, not merely a good in-sample reconstruction of returns.

Among the methods we compare, none satisfies all three simultaneously.
PCA and Sparse PCA satisfy P3 to some degree \citep{fan2013} but fail P1 (no link to characteristics)
and P2 (factors are statistical, not economic).
IPCA satisfies P1 by construction but remains linear, limiting both P2 and P3 at high
cross-sectional complexity.
GKX satisfies P1 only partially, since it does not rely on company characteristics alone,
and achieves strong return reconstruction, but is optimised for pricing
errors rather than covariance prediction, and the latent factors are not examined for economic interpretability.

\begin{figure}[t]
  \centering
  \includegraphics[width=\textwidth]{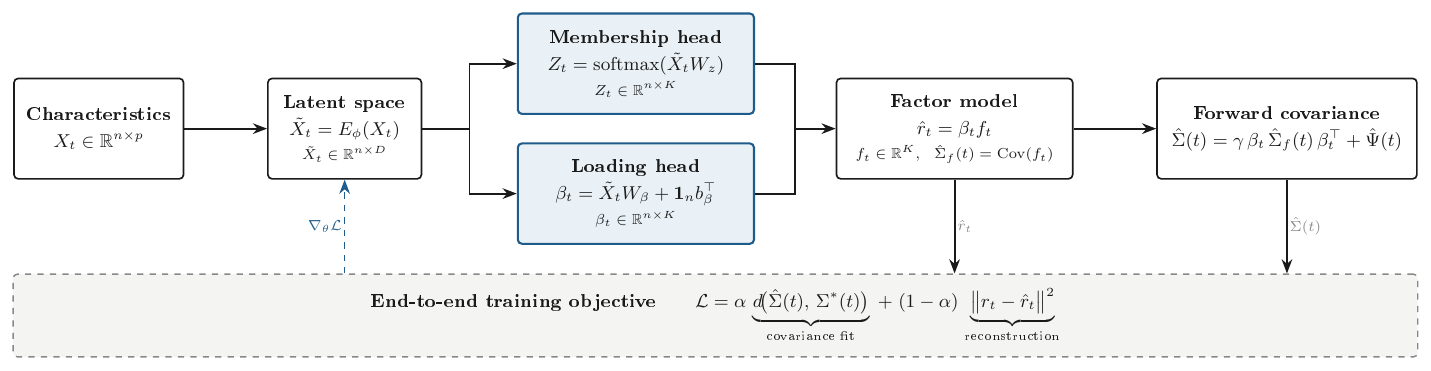}
  \caption{End-to-end architecture of CD-DFM. Firm characteristics $X_t$ are mapped by an encoder to a
    shared latent representation $\tilde{X}_t$. Two linear heads produce the soft factor memberships $Z_t$
    (constrained to the simplex) and the signed loadings $\beta_t$. The factor model reconstructs returns
    ($\hat r_t = \beta_t f_t$) and assembles the forward covariance $\hat\Sigma(t)$. The whole model is
    trained end-to-end by minimising a reconstruction term and a covariance-fit term
    against the realised forward covariance $\Sigma^*(t)$, and the gradient $\nabla_\theta\mathcal{L}$
    propagates back to the encoder.}
  \label{fig:architecture}
\end{figure}

We propose the \textbf{Characteristic-Driven Dynamic Factor Model} (CD-DFM), which satisfies all three properties simultaneously.
Each asset is assigned a soft membership vector $z_{i,t} \in \Delta^{K}$ over $K$ latent factors, computed
as a non-linear function of its observable characteristics. This drives an economically structured
cross-sectional latent space, from which we derive interpretable factors and a competitive covariance estimate. Figure~\ref{fig:architecture} sketches the full pipeline, from firm characteristics through the shared
latent representation and its two factor heads to the forward covariance estimate, and is detailed in
Section~\ref{sec:methodology}.

The characteristics are fundamentals such as size, valuation, and leverage
(Section~\ref{sec:data}), not the reactive, return-based signals most factor models rely on. They barely
move from day to day, so the model sees far fewer genuinely distinct observations than a return series
would provide, and yet it estimates risk competitively from them. Returns are
not part of the encoder inputs. The latent geometry and the loadings are functions of the
characteristics alone, and returns enter only to realise the factors and to form the covariance target,
in window-averaged form.

The model is evaluated on a panel of S\&P 500 constituents ($n = 239$ stocks) using daily returns, with an out-of-sample
evaluation period covering 2022--2024.
The resulting factors capture both common market exposures and cross-sectional sources of variation, reflecting the relative
differences between firms beyond aggregate market movements.
Unlike PCA-based approaches, CD-DFM naturally accommodates previously unseen assets through their observable
characteristics alone, without requiring past return history.
The model is trained end-to-end on a risk-oriented objective that pairs a factor reconstruction with a
forward Stein loss, directly targeting the out-of-sample covariance.
We show empirically that observable characteristics alone are sufficient to construct a stable economic latent geometry
of the equity cross-section, from which CD-DFM derives well-calibrated covariance estimates, interpretable sector-themed
factors stable across market regimes, and reliable onboarding of unseen assets. We do not claim to outperform the
state-of-the-art on each task independently, but rather that CD-DFM stays competitive across all of them
at once, which no other benchmarked method does.

\section{A Characteristic-Driven Latent Factor Representation}

\subsection{Problem Formulation}

Let $r_t = (r_{1,t}, \ldots, r_{n,t})^\top \in \mathbb{R}^n$ be the vector of
daily log-returns of $n$ assets at time $t = 1, \ldots, T$.
For each asset $i$ and date $t$, we observe a vector of $p$ predetermined
characteristics $x_{i,t} \in \mathbb{R}^p$, collected in the matrix
$X_t \in \mathbb{R}^{n \times p}$.
Characteristics are available at the start of period $t$ and carry no
look-ahead information about $r_t$.

We seek a mapping of the form
\begin{equation}
  X_t \;\longmapsto\; \tilde{X}_t \;\longmapsto\; \bigl(Z_t,\, \beta_t,\, f_t,\, \hat{\Sigma}(t)\bigr),
\end{equation}
where $\tilde{X}_t \in \mathbb{R}^{n \times D}$ is a latent representation of
the cross-section, $Z_t \in \mathbb{R}^{n \times K}$ are soft factor memberships,
$\beta_t \in \mathbb{R}^{n \times K}$ are asset loadings, $f_t \in \mathbb{R}^K$
are factor returns, and $\hat{\Sigma}(t) \in \mathbb{R}^{n \times n}$ is a
covariance estimate.
Critically, returns $r_t$ enter only in the construction of $f_t$ and in the
training loss, but the latent representation $\tilde{X}_t$ is built solely from the
characteristics $X_t$. This is what makes it possible to embed a previously unseen asset
without any return history. The mapping is required to satisfy three objectives simultaneously:
\begin{enumerate}[label=(\roman*)]
  \item $\tilde{X}_t$ organises the cross-section into an economically
    meaningful latent geometry, so that assets with similar characteristics
    occupy nearby positions;
  \item $\beta_t$ and $f_t$ yield interpretable, sector-themed factors whose
    identities are stable over time;
  \item $\hat{\Sigma}(t)$ is a well-calibrated predictor of the forward
    realised covariance $\Sigma^*(t) = \frac{1}{\Delta}\sum_{s=t+1}^{t+\Delta} r_s r_s^\top, \, \Delta \gg 1$.
\end{enumerate}

These correspond to properties P1, P2, and P3 stated in the introduction.

\subsection{Methodology}
\label{sec:methodology}

This section details the architecture summarised in Figure~\ref{fig:architecture}: the characteristic
encoder, the two factor heads (memberships and loadings), the factor model, and the resulting forward
covariance, together with the end-to-end training objective.

\subsubsection{Characteristic encoding}

Raw characteristics $x_{i,t} \in \mathbb{R}^p$ are heterogeneous, some are
continuous (market capitalisation, leverage, revenue, ...), others are categorical
(sector, industry). First, they are normalised \emph{cross-sectionally}
at each date, so each feature describes an asset's position relative to its peers at
time $t$, not its absolute level. These two types of features require different treatment.

On one hand, each continuous scalar
$x_{i,t,j} \in \mathbb{R}$ is mapped into a learned vector $v_{i,t,j} \in \mathbb{R}^d$
via a feature-specific affine map: $v_{i,t,j} = x_{i,t,j} \cdot w_j + b_j$,
where $w_j, b_j \in \mathbb{R}^d$ are trainable \citep{gorishniy2021}.
This gives each continuous feature its own $d$-dimensional token, allowing the
network to modulate the contribution of each feature independently of its scale.

On the other hand, each categorical feature (e.g.\ sector with $C$ categories) is mapped through a
dedicated embedding table $\mathrm{Emb}: \{1,\ldots,C\} \to \mathbb{R}^d$.
Categorical variables carry no natural ordering, and a lookup table avoids imposing
one.

The resulting tokens and categorical embeddings are concatenated into a single
vector and passed through a shallow MLP, producing the latent embedding of each asset $i$:
\begin{equation}
  \tilde{x}_{i,t} = E_\phi(x_{i,t}) \in \mathbb{R}^D, \qquad %
  \tilde{X}_t := \bigl(\tilde{x}_{1,t}, \ldots, \tilde{x}_{n,t}\bigr)^\top \in \mathbb{R}^{n \times D},
\end{equation}
stacking the per-asset embeddings as the rows of $\tilde{X}_t$, exactly as $X_t$ stacks the raw
characteristics $x_{i,t}$.

The encoder $E_\phi$ is applied independently to each asset and date, so $\tilde{X}_t$
depends only on $X_t$, never on $r_t$.
For an asset with characteristics $x\in\mathbb{R}^p$, the embedding $E_\phi(x)$ is a
$D$-dimensional vector encoding a nonlinear transformation of its fundamentals, which one may read as a
\emph{nonlinear factor} representation of the asset: a compression of the characteristics when $D<p$, and
an expansion when $D>p$. In this sense,
the dimension $D$ controls the expressivity
of the characteristic representation. Even if the manifold hypothesis says that the meaningful variation
of our financial assets lives on a low-dimensional submanifold of $\mathbb{R}^p$, this manifold may have
complex geometry. A larger $D$ gives the encoder enough room to ``unfold'' this
geometry into a more regular Euclidean space, and allows the encoder to capture non-linear interactions (e.g.,
``high news flow \emph{and} negative sentiment \emph{and} financial sector'').

\subsubsection{Soft memberships, factor returns, and loadings}

Given the latent representation $\tilde{X}_t \in \mathbb{R}^{n \times D}$, two parallel linear
heads derive the factor structure: one assigns assets to factors, the other converts those
assignments into portfolios and signed exposures.

\paragraph{Obtaining soft memberships.}
A projection $W_z \in \mathbb{R}^{D \times K}$ maps each asset's latent embedding to a
$K$-dimensional score, and a row-wise softmax turns the scores
into a membership matrix
\begin{equation}
  Z_t = \mathrm{softmax}\!\left(\tilde{X}_t W_z  \right) \in \mathbb{R}^{n \times K},
  \qquad Z_{i,k,t} > 0, \quad \textstyle\sum_k Z_{i,k,t} = 1.
\end{equation}
Each row $z_{i,t} \in \Delta^{K}$ is the soft probability that asset $i$ belongs to
nonlinear factor $k$ at date $t$. Because it sums to one, it is a convex decomposition of the asset over
this nonlinear factor basis, and assets with similar characteristics thus receive similar membership
vectors.

\paragraph{Computing nonlinear factor returns.}
The most direct way to turn memberships into factor returns is to form, for each factor $k$,
the membership-weighted cross-sectional average $f_{k,t} = w_{k,t}^\top r_t$ with
$w_{k,t} = Z_{\cdot,k,t}/\sum_i Z_{i,k,t} \geq 0$, that is, a long-only portfolio of the assets
most strongly assigned to factor $k$. This construction is structurally problematic, because
all $K$ portfolios are long-only combinations of the same cross-section, so every one of
them loads heavily on the common market mode and membership alone cannot separate their return
dynamics. Hence, the resulting factors cannot decorrelate from one another.

We instead isolate the market explicitly and use the remaining memberships to build
cash-neutral residual factors. The first factor is the equal-weight market return,
independent of $Z_t$,
\begin{equation}
  f_{0,t} = \frac{1}{n}\sum_{i=1}^n r_{i,t}.
\end{equation}
The remaining $K-1$ factors are long-short portfolios built from the centred membership
columns $k=1,\dots,K-1$, rescaled to unit gross exposure,
\begin{equation}
  f_{k,t} = w_{k,t}^\top r_t, \qquad
  w_{k,t} = \frac{z_{k,t} - \bar z_{k,t}\mathbf{1}_n}{\lVert z_{k,t} - \bar z_{k,t}\mathbf{1}_n \rVert_1}, \quad
  \bar z_{k,t} = \frac1n \sum_i Z_{i,k,t}, \qquad k = 1,\dots,K-1,
\end{equation}
so that $\mathbf{1}_n^\top (z_{k,t} - \bar z_{k,t}\mathbf{1}_n) = 0$ and $\lVert w_{k,t}\rVert_1 = 1$.
Each residual factor is thus cash-neutral (dollar-neutral, hence purely cross-sectional)
by construction and orthogonal to the market mode,
free to carry a distinct source of cross-sectional variation, and of comparable scale across
factors and dates. Stacking $f_{0,t}$ and $(f_{k,t})_{k\geq1}$ gives $f_t \in \mathbb{R}^K$.
Crucially, $f_t$ depends on $r_t$ only through the memberships $Z_t$ and the fixed equal
weights of $f_{0,t}$, and both are determined by $X_t$ alone.
As a consequence, a new, unseen asset can be added to the model without any retraining and
without any return history: it is assigned to the factor basis directly from its characteristics.

\paragraph{Estimating loadings on the nonlinear factors.}
Memberships constrain each asset to a non-negative exposure to each factor, reflecting a
long-only portfolio construction. To allow signed exposures and a richer decomposition of
returns, a separate linear head on the same latent embedding produces the loading matrix
\begin{equation}
  \beta_t = \tilde{X}_t W_\beta + \mathbf{1}_n b_\beta^\top \in \mathbb{R}^{n \times K},
\end{equation}
with $W_\beta \in \mathbb{R}^{D \times K}$ and bias $b_\beta \in \mathbb{R}^K$. Unlike $Z_t$,
the loadings $\beta_t$ are unconstrained in sign, so each asset can hold both long and short
exposures to each factor, and the return of asset $i$ is reconstructed as
$\hat{r}_{i,t} = \beta_{i,t}^\top f_t$. The two heads $W_z$ and $W_\beta$, together with the
encoder $E_\phi$, are learned jointly, end-to-end.

\subsubsection{Covariance estimation}

The loadings and factor returns combine into a covariance estimate built from three pieces, a
factor covariance, an idiosyncratic diagonal, and a single scale correction. The factor covariance
$\hat{\Sigma}_f(t) \in \mathbb{R}^{K \times K}$ is estimated on a trailing lookback window
$[t-H+1,\,t]$ by an exponentially weighted moving average of the factor returns over that window:
\begin{equation}\label{eq:sigmaf}
  \hat{\Sigma}_f(t) = \sum_{s = t-H+1}^{t} w_{t-s}\, f_s f_s^\top, \qquad
  w_{t-s} \propto \lambda^{\,t-s}, \quad \textstyle\sum_{s} w_{t-s} = 1,
\end{equation}
with $\lambda \in (0,1)$ chosen so that the weights decay over a fixed half-life.
This is a rolling estimate of $\mathbb{E}\!\left[f_t f_t^\top\right]$, the covariance of the nonlinear
factors, with a weighting scheme that accommodates non-stationarity.
Recent
observations carry more weight, so
$\hat{\Sigma}_f(t)$ tracks the recent
regime rather than
averaging uniformly over the full lookback. The chosen
half-life is given in the appendix.
The idiosyncratic variances come from the reconstruction residuals
$e_{i,s} = r_{i,s} - \beta_{i,s}^\top f_s$ accumulated over the same window,
\begin{equation}\label{eq:psi}
  \hat{\Psi}(t) = \mathrm{diag}\Bigl(\max\bigl(\overline{e_i^2}(t),\, \psi_{\min}\bigr)\Bigr)_{i=1}^n,
  \qquad \overline{e_i^2}(t) = \frac1H \sum_{s=t-H+1}^{t} e_{i,s}^2,
\end{equation}
floored at a small $\psi_{\min} > 0$ so that $\hat{\Psi}(t)$
stays strictly positive definite.
The two pieces assemble into a natural estimate of the return second moment $\mathbb{E}\!\left[r_t r_t^\top\right]$,
\begin{equation}\label{eq:lowrank}
  \hat{\Sigma}(t) = \gamma\, \beta_t\, \hat{\Sigma}_f(t)\, \beta_t^\top + \hat{\Psi}(t),
\end{equation}
i.e.\ the canonical low-rank-plus-diagonal form of a factor-model covariance. This structure is what
keeps $\hat{\Sigma}(t)$ well-conditioned and invertible even when the number of assets $n$
exceeds the length of the estimation window, a regime in which the sample covariance of returns
is singular. The rank of the systematic block is left at the full factor dimension $K$ rather than
truncated or shrunk further, because $K$ is already small relative to $n$.
The scalar $\gamma > 0$ is learned jointly with the rest of the model and acts as an amplitude
dial. Empirical factor covariances are known to understate risk, and our factors are a projection of the returns, so they carry less variance which lead the systematic block
$\beta_t\,\hat{\Sigma}_f(t)\,\beta_t^\top$ to come out too small. Learning $\gamma$ rescales it back to
the right magnitude, leaving the geometry fixed by $\beta_t$ and $\hat{\Sigma}_f(t)$ untouched.

\subsubsection{Training objective}

The model is trained end to end on two terms, a forward covariance loss and a factor reconstruction
loss. The covariance loss carries the objective and targets the risk of the cross-section directly,
while the reconstruction keeps the factors tied to the returns. Everything the encoder produces, the
latent geometry, the memberships, the loadings, and the scale $\gamma$, is fit jointly by minimising
their weighted sum.

The covariance term drives the objective. We want $\hat{\Sigma}(t)$ to match the covariance of the
future returns, since a covariance that fits the past but not the future is useless for risk.
We measure the gap between the model estimate and the realised future with
the Stein loss \cite{james1961estimation,demiguel2009generalized}:
\begin{equation}
  d\bigl(\hat{\Sigma}(t),\, \Sigma^*(t)\bigr)
    = \operatorname{Tr}\!\bigl(\Sigma^*(t)\,\hat{\Sigma}(t)^{-1}\bigr)
    - \log\det\!\bigl(\Sigma^*(t)\,\hat{\Sigma}(t)^{-1}\bigr) - n ,
\end{equation}
where $\Sigma^*(t) = \frac1\Delta \sum_{s=t+1}^{t+\Delta} r_s r_s^\top$ is the realised future
second-moment matrix. Since $\Sigma^*(t)$ does not depend on $\theta$, the $\log\det \Sigma^*(t)$ term
is constant in the optimisation. Dropping constants, the training loss at timestep $t$ reduces to
\begin{equation}\label{eq:lcov}
  \mathcal{L}_{\mathrm{cov}}(t) = \log\det \hat{\Sigma}(t)
     + \frac1\Delta \sum_{s=t+1}^{t+\Delta} r_s^\top \hat{\Sigma}(t)^{-1} r_s .
\end{equation}

This distance is not an arbitrary choice. For two zero-mean Gaussians it is exactly twice their KL
divergence,
\begin{equation}
  \mathrm{KL}\bigl(\mathcal{N}(0,\Sigma^*) \,\Vert\, \mathcal{N}(0,\hat{\Sigma})\bigr)
    = \tfrac12\, d\bigl(\hat{\Sigma},\Sigma^*\bigr),
\end{equation}
so minimising $\mathcal{L}_{\mathrm{cov}}$ pulls the return distribution implied by $\hat{\Sigma}(t)$
towards the one that generated the future returns. We are matching distributions, not matrix entries.

This is also why we use Stein rather than a plain Frobenius error on $\hat{\Sigma}(t)$, as it
scores the risk geometry. The $\log$ determinant and the inverse quadratic term blow up when
$\hat{\Sigma}(t)$ is near-singular or has a badly estimated small eigenvalue, so the loss punishes
non-SPD matrices and mispriced low-variance directions far harder than large ones. Those small
eigenvalues are exactly what portfolio construction is most sensitive to, since the optimiser inverts
the covariance \cite[Section 2.2]{cetingoz2026synthetic}.
Training on Stein therefore forces a coherent risk geometry, instead of a matrix that
merely looks close entry by entry.

The distance uses $\hat{\Sigma}(t)^{-1}$ but never $\Sigma^{*}(t)^{-1}$, which matters here. The
realised $\Sigma^*(t)$ is a sum of $\Delta$ rank-one terms and is singular whenever $\Delta < n$, so
it cannot be inverted. The model estimate is not as the low-rank-plus-diagonal form \eqref{eq:lowrank}
stays SPD and invertible by construction, whatever the number of assets. The loss only ever inverts
the covariance estimated by the model, so the singularity of the target is not a problem.

The second term is a reconstruction that anchors the factor structure to the returns, asking the
factor model to reproduce each in-sample return,
\begin{equation}\label{eq:lrec}
  \mathcal{L}_{\mathrm{rec}}(t) = \frac1{H}\sum_{s=t-H+1}^{t}
     \bigl\lVert r_s - \beta_s f_s \bigr\rVert^2 .
\end{equation}
It keeps $\beta_t$ and $f_t$ tied to real return variation and stabilises training, otherwise the
Stein term alone could reach a good likelihood with factors that no longer explain the returns.

The full objective is a convex combination of the two,
\begin{equation}\label{eq:ltot}
  \mathcal{L}(t) = \alpha\,\mathcal{L}_{\mathrm{cov}}(t) + (1-\alpha)\,\mathcal{L}_{\mathrm{rec}}(t),
  \qquad \alpha \in (0,1),
\end{equation}
averaged over the training anchors $t$, and we tilt $\alpha$ towards the covariance term.

\subsection{Training algorithm}

The whole pipeline is differentiable, so every parameter, the encoder, the two projection heads, and
the scale $\gamma$, is trained together by gradient descent on $\mathcal{L}$. Training runs over
timesteps sampled on a fixed stride through the training period. At each timestep the model reads the
characteristics, builds the factor structure and the covariance $\hat{\Sigma}(t)$ from the lookback
window, and is scored against the forward window. Timesteps are processed in mini-batches with Adam,
and we early-stop on the validation Stein loss. Algorithm~\ref{alg:train} states the loop.

Evaluating $\mathcal{L}_{\mathrm{cov}}$ needs $\hat{\Sigma}(t)^{-1}$ and $\log\det\hat{\Sigma}(t)$,
which look like $n \times n$ operations. The low-rank-plus-diagonal form \eqref{eq:lowrank} lets us
compute both through a $K \times K$ system instead, at a cost linear in $n$. Appendix~\ref{app:woodbury}
gives the
derivation.

\begin{algorithm}[H]
  \caption{CD-DFM training}
  \label{alg:train}
  \begin{algorithmic}[1]
    \Require returns $\{r_t\}$, characteristics $\{X_t\}$; lookback $H$, horizon $\Delta$, weight $\alpha$, stride $\sigma$
    \State Initialise parameters $\theta$;\ \ $\theta^\star \gets \theta$
    \State Sample train / val timesteps every $\sigma$ days from $\{H, \ldots, T-\Delta\}$
    \For{each epoch, until the validation loss stops improving}
      \For{each mini-batch of timesteps $\mathcal{T}$}
        \For{each timestep $t \in \mathcal{T}$}
          \State $\tilde{X}_t \gets E_\phi(X_t)$;\quad $Z_t \gets \mathrm{softmax}(\tilde{X}_t W_z)$;\quad $\beta_t \gets \tilde{X}_t W_\beta$
          \State $f_s \gets$ factor returns from $(Z_t, r_s)$
          \State Compute $\hat{\Sigma}(t)$ from $\hat{\Sigma}_f(t)$, $\hat{\Psi}(t)$, $\gamma$ \Comment{Eqs.~\eqref{eq:sigmaf}, \eqref{eq:psi}, \eqref{eq:lowrank}}
          \State $\mathcal{L}(t) \gets \alpha\,\mathcal{L}_{\mathrm{cov}}(t) + (1-\alpha)\,\mathcal{L}_{\mathrm{rec}}(t)$ \Comment{Eqs.~\eqref{eq:lcov}, \eqref{eq:lrec}, \eqref{eq:ltot}}
        \EndFor
        \State $\theta \gets \mathrm{Adam}\bigl(\theta,\ \nabla_\theta \tfrac1{|\mathcal{T}|}\sum_{t\in\mathcal{T}} \mathcal{L}(t)\bigr)$
      \EndFor
      \State Score the validation Stein loss;\ \ if it improves, $\theta^\star \gets \theta$
    \EndFor
    \State \Return $\theta^\star$
  \end{algorithmic}
\end{algorithm}

\section{Data}
\label{sec:data}

The latent structure is driven entirely by the
characteristics $x_{i,t}$. They decide where an asset sits in the factor latent space,
and therefore how its
membership weights $Z_{i,k,t}$ are assigned. This constrains what the features should describe: what an
asset \emph{is} economically, not how its price returns have historically behaved.

Moreover, that gives the model a natural interchangeability property: two assets
with identical characteristics get the same memberships and land at the same point in latent space.
The practical payoff is that when a stock leaves the universe and another takes its place, the newcomer's
memberships follow directly from its characteristics, with no retraining.

Relying only on fundamentals means the latent factors capture the joint moves of asset prices from
static characteristics alone.
This rules out anything derived from prices or trading activity, momentum, realised volatility, market
beta, since those move with market conditions and break the link between economic identity and latent
position. A purely static feature set has one blind spot, though: it cannot capture the sharp rise in
cross-sectional correlations during stress. We therefore add a few macro-reactive signals, news flow
and commodity dependence, which let the encoder deform the membership matrix $Z_t$ across regimes
without breaking interchangeability. Table~\ref{tab:features} lists the full set; all features are
standardised cross-sectionally at each date.

\paragraph{Universe and features.} We work with a fixed universe of $239$ S\&P~500 stocks, present on
every date, over 2011--2024 ($3521$ daily observations per stock). Each stock--day carries $22$
predetermined characteristics: $10$ continuous (e.g.\ size, book-to-market, leverage), $10$
compositional (geographic and commodity revenue shares), and $2$ categorical (GICS sector and
exchange). No market data enters the features: no price, no return, no realised volatility. Returns are
used only to form the factors and the training loss.

\begin{table}[htbp]
\centering
\small
\begin{tabular}{lll}
\toprule
\textbf{Feature} & \textbf{Description} & \textbf{Update frequency} \\
\midrule
\multicolumn{3}{l}{\textit{Structural}} \\
GICS sector          & Industry classification (11 sectors)          & Rare \\
Exchange             & Listing venue                                 & Rare \\
Geographic exposure  & Revenue shares by region (5 shares)           & Annual \\
Commodity dependence & Revenue shares by commodity (5 shares)         & Annual \\
\midrule
\multicolumn{3}{l}{\textit{Fundamental}} \\
Log market cap       & Firm size                                     & Daily \\
Book-to-market       & Value vs.\ growth                             & Quarterly \\
Leverage             & Indebtedness                                  & Quarterly \\
Return on equity     & Profitability                                 & Quarterly \\
Capex / revenue      & Capital intensity                             & Quarterly \\
Working capital      & Short-term operating position                 & Quarterly \\
P/E proxy            & Valuation / duration proxy                    & Quarterly \\
\midrule
\multicolumn{3}{l}{\textit{Macro-reactive}} \\
News flow            & Daily volume of news on the asset             & Daily \\
News sentiment       & Tone of that news                             & Daily \\
News buzz            & Abnormal attention                            & Daily \\
\bottomrule
\end{tabular}
\caption{Input characteristics used to build the latent representation. The structural block captures
  economic identity and enforces interchangeability; the macro-reactive block lets the model respond to
  regime and stress episodes. Most features update quarterly or less, so the panel is effectively low
  frequency despite the daily sampling.}
\label{tab:features}
\end{table}

\paragraph{Data sources.} The financial and structural characteristics, fundamentals, revenue
shares, GICS classification, and the prices used to form returns, are drawn from \emph{S\&P Global
Compustat Xpressfeed}, with companies keyed by their Compustat identifier (\texttt{gvkey}). The three
news features come from \emph{RavenPack Edge} company-level analytics, using only information published
strictly before each date. The panel spans 2011--2024, starting in 2011 because per-company news coverage
is unavailable before then.

\begin{figure}[htbp]
\centering
\includegraphics[width=\textwidth]{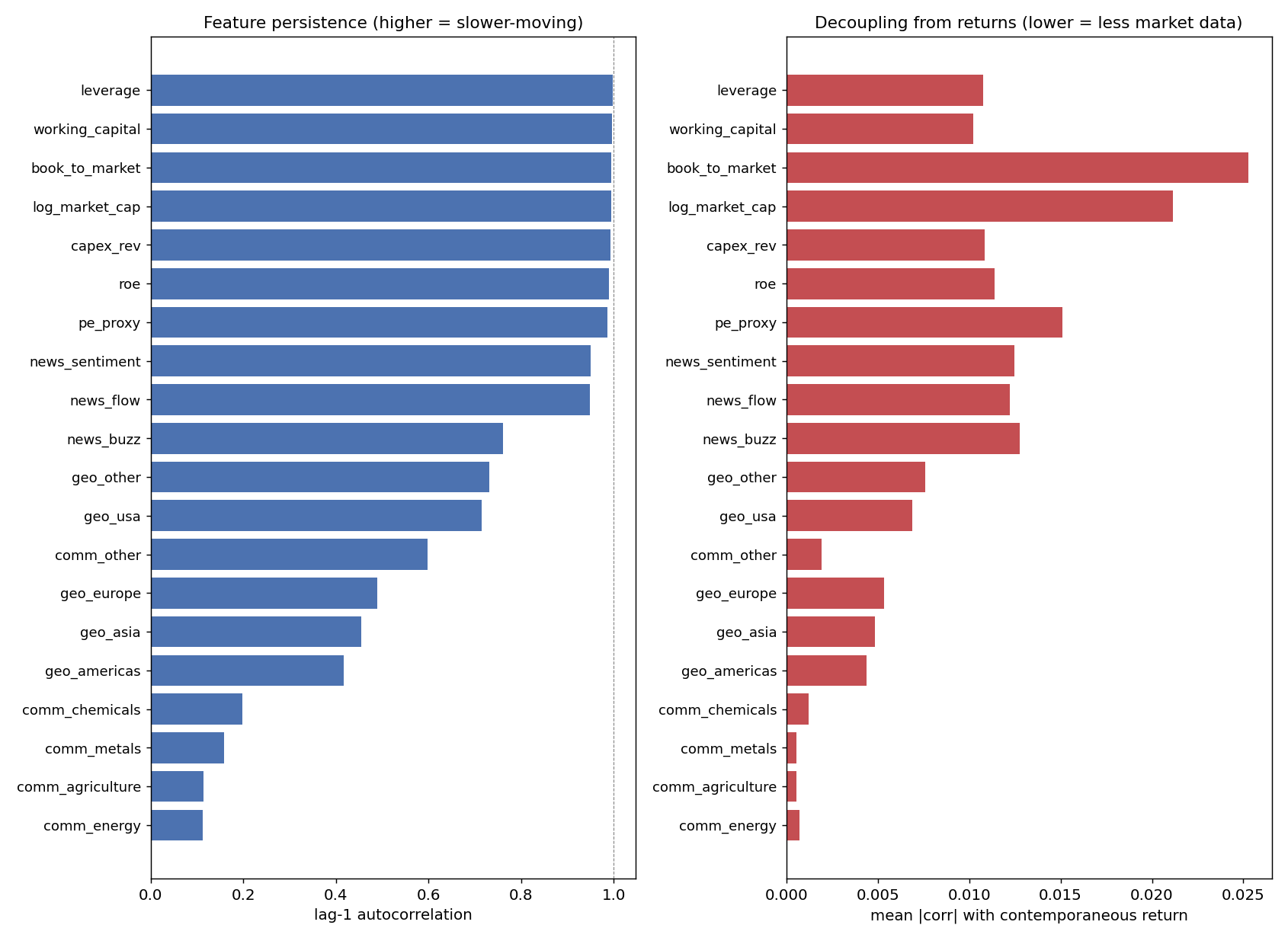}
\caption{Feature statistics on the processed panel. Left: lag-1 autocorrelation, the features are
  highly persistent (fundamentals near $1$). Right: mean absolute correlation between each feature and
  the same-day return, near zero for every feature, confirming the inputs carry no direct market
  information.}
\label{fig:dataset}
\end{figure}

\paragraph{A very low-frequency panel.} The observations are daily, but the information in them is not.
Most characteristics refresh on a quarterly or annual cycle, and even the fastest-moving ones carry
little day-to-day variation. Two properties make the task deliberately hard. First, the features are
slow-moving. The mean lag-1 autocorrelation is $0.68$, with accounting fundamentals near $1$ since they
only update quarterly, and the cross-sectional variance is on average $20$ times the within-stock
temporal variance. Each stock has an almost static fingerprint rather than a live feed. Second, they
are decoupled from returns: the mean absolute correlation between a feature and the same-day return is $0.009$.
Recovering a forward return-covariance structure from inputs that individually carry almost no return
information is a genuine test, not a repackaging of past prices. Table~\ref{tab:dataset} and
Figure~\ref{fig:dataset} give the breakdown.

\begin{table}[htbp]
\centering
\small
\begin{tabular}{lccccc}
\toprule
Feature group & \# & ACF$_1$ & staticness & change rate & $|\mathrm{corr}(\cdot,r)|$ \\
\midrule
Fundamentals & 7 & 0.99 & 2.7 & 0.24 & 0.015 \\
News & 3 & 0.89 & 0.9 & 0.85 & 0.013 \\
Geography & 5 & 0.56 & 7.5 & 0.00 & 0.006 \\
Commodities & 5 & 0.24 & 69.5 & 0.00 & 0.001 \\
\midrule
All (numeric) & 20 & 0.68 & 20.3 & 0.21 & 0.009 \\
\bottomrule
\end{tabular}
\caption{Feature statistics by group. ACF$_1$: mean lag-1 autocorrelation (persistence).
  staticness: variation across stocks divided by variation over time (high means it separates firms but
  barely moves through time). Change rate: fraction of days the
  value moves. $|\mathrm{corr}(\cdot,r)|$: mean absolute correlation between the feature and the
  same-day return. Fundamentals are near-static; commodity and geography shares vary almost only across
  stocks; every group is essentially uncorrelated with returns.}
\label{tab:dataset}
\end{table}

The universe is $N=239$ stocks and $K=11$ factors, selected by a Mar\v{c}enko--Pastur criterion on the
factor return covariance. The panel is split chronologically into train, validation, and a held-out
test period; unless stated otherwise, every reported number is computed strictly out-of-sample on the
2022--2024 test period.

\section{Experiments}
\label{sec:experiments}

We evaluate CD-DFM along the three properties it is designed for, a characteristic-driven latent
representation (P1), interpretable factors (P2), and a well-calibrated forward covariance (P3). Each
gets its own subsection, and we bring in competing models where a comparison is meaningful. Unless
noted, results are reported as a $5$-seed mean, strictly out-of-sample on
2022--2024.

\subsection{Latent representation}
\label{sec:latent}

The first property is that the latent space should be organised by economic identity. Stocks with
similar characteristics should sit close together, and the geometry should line up with something
recognisable. We measure this against GICS sectors and against the return cross-section, and we compare
to the autoencoder asset-pricing model of \citet{gu2019} (GKX), the one competitor that also
learns a latent embedding from firm characteristics, though ones mixed with market-driven metrics.

\paragraph{Metrics.} Write $\tilde{x}_i \in \mathbb{R}^D$ for the latent embedding of stock $i$,
averaged over the test dates ($\tilde{x}_i = \frac1{|\mathcal{T}|}\sum_{t\in\mathcal{T}}\tilde{x}_{i,t}$)
so that each stock has a single position, and
$d(i,j) = \lVert \tilde{x}_i - \tilde{x}_j\rVert$ for the latent distance. Let $s(i)$ be the GICS
sector of $i$. We define five metrics that probe the latent geometry from different angles.

\begin{itemize}[leftmargin=1.4em, topsep=2pt, itemsep=2pt, parsep=0pt]
  \item \emph{Sector silhouette}, the silhouette score of the embedding against sector labels. For
    stock $i$, let $\bar d\big(i, s(i)\big)$
    be its mean distance to the other stocks in its own sector and $\bar d\big(i, \neg s(i)\big)$
    its mean
    distance to the stocks of the nearest other sector; the silhouette is
    \begin{equation}
      \frac1N \sum_{i} \frac{\bar d\big(i, \neg s(i)\big) - \bar d\big(i, s(i)\big)}{\max\big(\bar d(i, \neg s(i)),\, \bar d(i, s(i))\big)} \in [-1, 1] .
    \end{equation}
    Higher means tighter, better-separated sector clusters.

  \item \emph{Separation ratio}, the ratio of mean inter-sector to mean intra-sector distance,
    \begin{equation}
      \frac{\operatorname{mean}\{ d(i,j) : s(i) \neq s(j) \}}
           {\operatorname{mean}\{ d(i,j) : s(i) = s(j) \}} .
    \end{equation}
    A value above $1$ means stocks from different sectors sit further apart than stocks from the same
    sector; higher is better.

  \item \emph{Separation AUC}, the area under the ROC curve of a classifier that ranks stock pairs by
    latent similarity $-d(i,j)$ and predicts whether they share a sector. It asks whether latent proximity
    alone recovers sector membership. $0.5$ is random, $1$ is perfect.

  \item \emph{Far-tail ordering}, the Spearman rank correlation \citep{spearman1904} between latent
    distance $d(i,j)$ and return de-correlation $1 - \mathrm{corr}(r_i, r_j)$, computed on the $10\%$
    most distant latent pairs only. It checks that
    the geometry stays ordered on the far tail, where dissimilar stocks should also be the least
    co-moving; higher is better.

  \item \emph{Neighbour overlap}, the Jaccard index \citep{jaccard1912} between each stock's $k$ nearest
    latent neighbours and its $k$ nearest return-correlation neighbours, averaged over stocks. It
	is the most direct probe of raw predictiveness, since it asks whether
    latent neighbours coincide with return neighbours. Higher is better.
\end{itemize}

\begin{figure}[htbp]
\centering
\begin{subfigure}{0.49\textwidth}
  \centering
  \includegraphics[width=\textwidth]{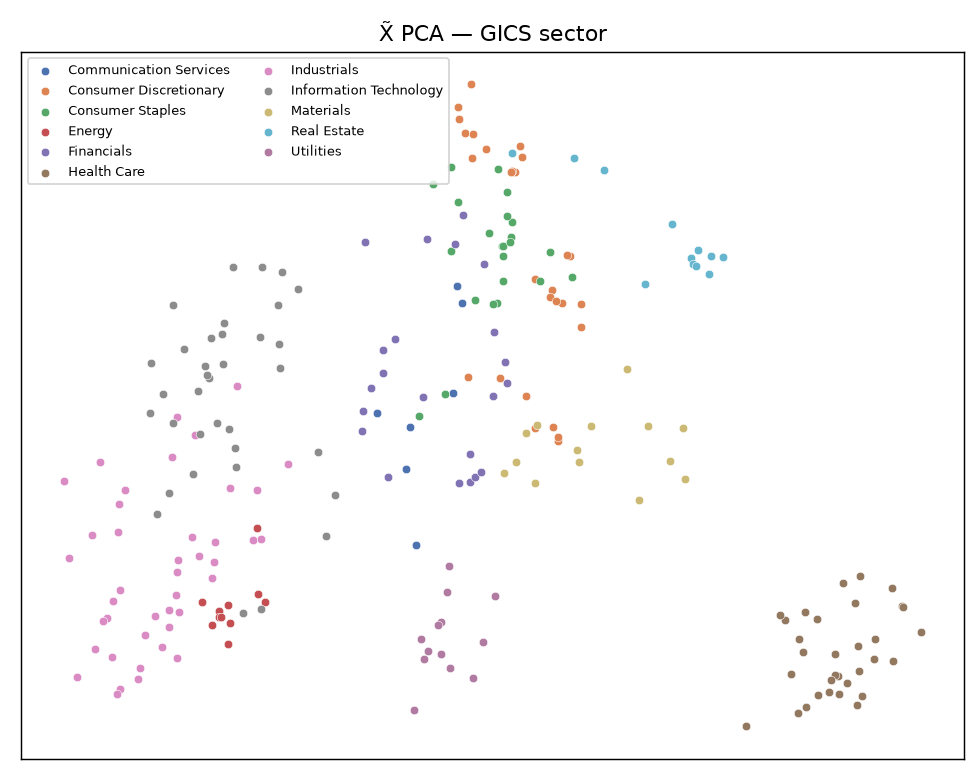}
  \caption{PCA}
\end{subfigure}
\hfill
\begin{subfigure}{0.49\textwidth}
  \centering
  \includegraphics[width=\textwidth]{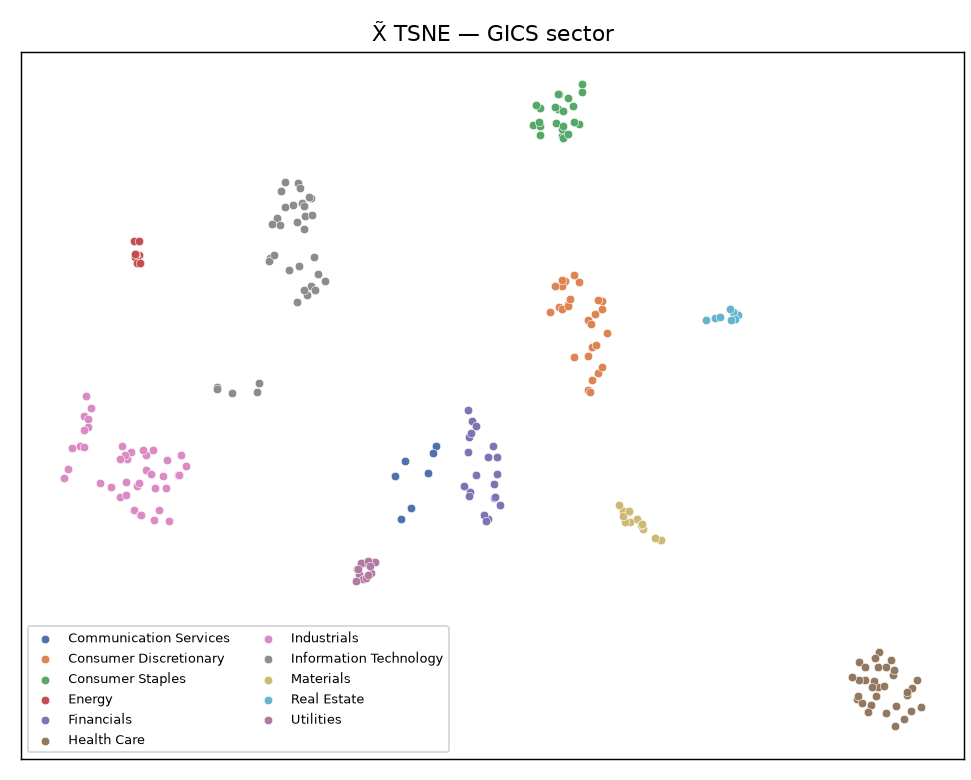}
  \caption{t-SNE}
\end{subfigure}
\caption{Two-dimensional views of the test-date-averaged latent embedding $\tilde{x}_i$, coloured by
GICS sector, obtained by PCA (left) and t-SNE (right).
Sectors form clean, well-separated clusters in both projections.}
\label{fig:latent_map}
\end{figure}

\paragraph{Results.} We compare against GKX \citep{gu2019}, the one benchmark with a comparable latent
embedding (Table~\ref{tab:latent_gkx}). CD-DFM produces the
more structured latent space. It is more sector-coherent, pushes dissimilar stocks further apart, and
stays weakly positive on the far tail, where GKX turns negative. GKX wins only on raw neighbour overlap. This
reflects a difference in how each model is anchored to returns. GKX optimises the pointwise
reconstruction of returns directly, so its nearest-neighbour graph aligns tightly with the
return-correlation graph, whereas CD-DFM targets the forward covariance through a latent geometry, which
trades some of that raw alignment for a more economically organised embedding. The sector organisation
of Figure~\ref{fig:latent_map} is consistent with this. Both maps separate the cross-section along
sector lines, and CD-DFM does so with tighter, better-separated clusters.

\begin{table}[htbp]
\centering
\small
\begin{tabular}{l C{2.2cm} C{2.2cm}}
\toprule
\textbf{Metric} & \textbf{CD-DFM} & \textbf{GKX} \\
\midrule
Sector silhouette   & \textbf{0.38} & 0.28 \\
Separation AUC      & \textbf{0.98} & \textbf{0.98} \\
Separation ratio    & \textbf{3.0} & 2.5 \\
Far-tail order      & \textbf{0.03} & $-0.02$ \\
Neighbour overlap   & 0.27 & \textbf{0.30} \\
\bottomrule
\end{tabular}
\caption{Latent-space comparison of CD-DFM and GKX average over 5 seeds. Higher is better for all
rows. Best in \textbf{bold}.}
\label{tab:latent_gkx}
\end{table}

\begin{figure}[htbp]
\centering
\includegraphics[width=0.72\textwidth]{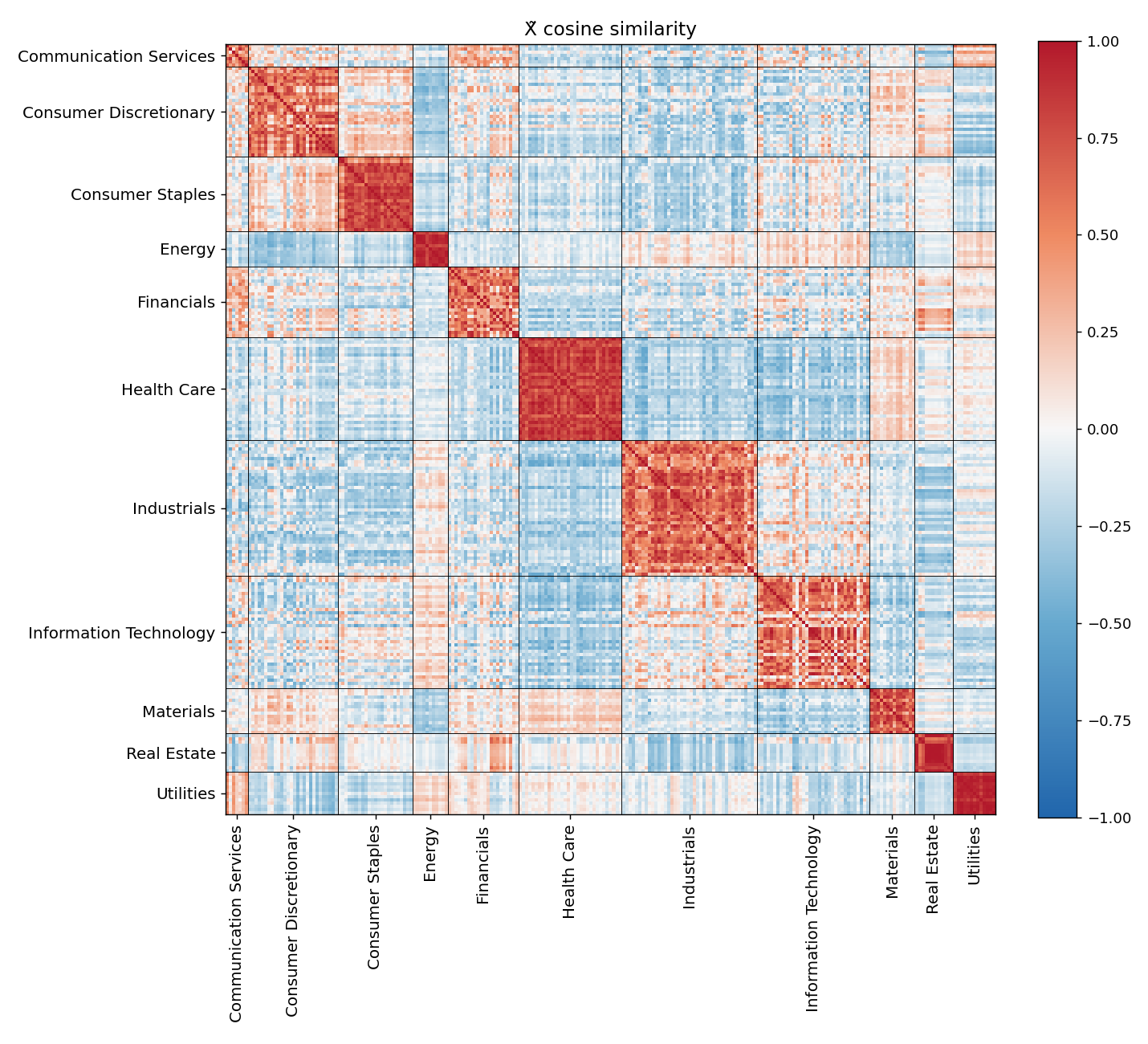}
\caption{Cosine similarity between the test-date-averaged latent embeddings $\tilde{x}_i$ (common mode
  removed), assets ordered by GICS sector. Red diagonal blocks show high same-sector similarity and
  off-diagonal pairs are near zero. The
  blocks are not uniform, and within a sector such as Information Technology firms with different
  fundamentals stay apart, so the geometry captures more than the sector label.}
\label{fig:cosine}
\end{figure}

\paragraph{Similarity structure.} Removing the common mode and taking the cosine similarity between
latent embeddings, ordered by sector, gives the block pattern of Figure~\ref{fig:cosine}. Same-sector
pairs are on average strongly positive along the diagonal blocks, and cross-sector pairs sit near zero
or negative. The sector is one of the input characteristics, so part of this alignment is expected.
What matters is that the geometry is not reducible to the sector label. The diagonal blocks are far from
uniform, and within a sector the model still pulls apart firms whose fundamentals differ. Information
Technology is the clearest case, as it spans large, cash-rich incumbents down to smaller, capital-hungry
growth names, and their latent similarity can be low despite a shared GICS code. For instance,
Apple and Juniper Networks are both Information Technology, yet their latent similarity is $-0.23$,
because they sit at opposite ends of the sector on size, valuation, leverage, and profitability. A
shared GICS code is not enough to make two firms close, which is what the non-uniform sector blocks in
Figure~\ref{fig:cosine} reflect.

\paragraph{Qualitative example.} To illustrate what a latent neighbourhood contains, take Microsoft and
list its nearest neighbours by latent cosine similarity (Table~\ref{tab:msft}). The three closest are
Adobe, Synopsys, and Nvidia, all Information Technology, all large, low book-to-market, US-centred
firms. The model groups Microsoft with software and semiconductor peers purely from characteristics,
which is the behaviour interchangeability is meant to produce.

\begin{table}[htbp]
\centering
\small
\begin{tabular}{lccccccc}
\toprule
Stock & cosine & log mkt cap & book/mkt & leverage & ROE & capex/rev & news flow \\
\midrule
MSFT & --- & 0.50 & $-0.28$ & $-0.20$ & 0.31 & 0.45 & 0.49 \\
\midrule
ADBE & 0.963 & 0.42 & $-0.30$ & $-0.27$ & 0.25 & 0.37 & 0.24 \\
SNPS & 0.924 & 0.19 & $-0.25$ & $-0.36$ & 0.01 & 0.17 & 0.03 \\
NVDA & 0.915 & 0.48 & $-0.36$ & $-0.18$ & 0.16 & 0.18 & 0.44 \\
\bottomrule
\end{tabular}
\caption{Microsoft and its three nearest latent neighbours, with a few characteristics for context
  (cross-sectional ranks). All three are large, low book-to-market, US-centred technology firms,
  matched from characteristics alone.}
\label{tab:msft}
\end{table}

\subsection{Interpretability}
\label{sec:interp}

The second property is that each factor should be an identifiable economic object, not an anonymous
statistical direction. We read the factors at three levels, their identity, what characteristics drive
them, and how they behave across market regimes.

\subsubsection{Factor identity}

Each factor gets a one-line identity, read from three views that should agree if the factor is genuinely
economic. Write $\overline{Z} \in \mathbb{R}^{n \times K}$ and $\overline{\beta} \in \mathbb{R}^{n
\times K}$ for the time-averaged memberships and loadings. The \emph{home sector} of factor $k$ is the
GICS sector on which it concentrates the most membership mass, $\arg\max_s \frac1{|s|}\sum_{i \in s}
\overline{Z}_{ik}$. Two concentration ratios then measure how localised the factor is on that home
sector $h$,
\begin{equation}
  \mathrm{ratio}_Z = \frac{\operatorname{mean}_{i \in h} \overline{Z}_{ik}}
                          {\operatorname{mean}_{i \notin h} \overline{Z}_{ik}}, \qquad
  \mathrm{ratio}_\beta = \frac{\operatorname{mean}_{i \in h} |\overline{\beta}_{ik}|}
                              {\operatorname{mean}_{i \notin h} |\overline{\beta}_{ik}|} .
\end{equation}
$\mathrm{ratio}_Z$ works on the memberships (who belongs to the factor), $\mathrm{ratio}_\beta$ on the
signed loadings (who is exposed to it): a value above $1$ means the factor sits mostly on its home
sector.

Table~\ref{tab:factor_id} reports these for all eleven factors. Every factor has $\mathrm{ratio}_Z > 1$
(median around $1.8$, up to $3.1$ for the energy factor), and ten of eleven also clear one on
$\mathrm{ratio}_\beta$. None is dead. Membership home, loading concentration, and the strongest
characteristic correlations line up factor by factor. Factor~6 is Energy with a positive
commodity-energy tilt, factor~7 Industrials with a metals tilt, factor~8 Utilities, factor~2 an
Asia-versus-US axis, factors~3 and~4 a value tilt. Factor~0 is the equal-weight market, carrying the bulk of the
return variance ($R^2 = 0.36$, an order of magnitude above the rest) and spreading its exposure evenly
($\mathrm{ratio}_\beta \simeq 1$), as a systematic factor should. The small non-market $R^2$ values
($0.007$ to $0.048$) are exactly what market-neutral long-short factors look like, since they carry
cross-sectional structure rather than market variance. Interpretability is judged on concentration and
coherence, not on variance share.

\newcommand{\upc}[1]{\textcolor{dkgreen}{$\uparrow$}\,#1}
\newcommand{\dnc}[1]{\textcolor{red}{$\downarrow$}\,#1}

\begin{table}[htbp]
\centering
\footnotesize
\setlength{\tabcolsep}{6pt}
\renewcommand{\arraystretch}{1.25}
\begin{tabular}{@{}c l ccc >{\raggedright\arraybackslash}p{5.4cm}@{}}
\toprule
Factor & Home sector & $\mathrm{ratio}_Z$ & $\mathrm{ratio}_\beta$ & $R^2$ & Top characteristics \\
\midrule
0  & Market (all)           & ---  & 1.00 & 0.364 & --- \\
1  & Real Estate            & 2.19 & 7.61 & 0.013 & \upc{capex/rev}\ \dnc{work.\ cap.}\ \dnc{news flow} \\
2  & Information Technology & 1.71 & 3.02 & 0.048 & \upc{geo.\ Asia}\ \dnc{geo.\ USA}\ \upc{comm.\ other} \\
3  & Real Estate            & 1.18 & 1.53 & 0.007 & \dnc{comm.\ energy}\ \upc{book/mkt}\ \upc{mkt cap} \\
4  & Financials             & 1.85 & 3.41 & 0.010 & \upc{book/mkt}\ \dnc{comm.\ agri.}\ \upc{comm.\ other} \\
5  & Health Care            & 1.79 & 3.95 & 0.025 & \upc{comm.\ chem.}\ \dnc{P/E}\ \upc{work.\ cap.} \\
6  & Energy                 & 3.06 & 5.76 & 0.035 & \upc{comm.\ energy}\ \dnc{comm.\ agri.}\ \dnc{P/E} \\
7  & Industrials            & 1.64 & 2.70 & 0.012 & \upc{comm.\ metals}\ \upc{P/E}\ \upc{ROE} \\
8  & Utilities              & 1.94 & 6.49 & 0.030 & \upc{comm.\ energy}\ \dnc{comm.\ other}\ \upc{book/mkt} \\
9  & Consumer Discretionary & 1.34 & 0.89 & 0.019 & \dnc{comm.\ agri.}\ \dnc{mkt cap}\ \dnc{ROE} \\
10 & Information Technology & 1.31 & 1.75 & 0.010 & \dnc{mkt cap}\ \upc{geo.\ Asia}\ \dnc{news flow} \\
\bottomrule
\end{tabular}
\caption{Factor identity over the whole test series. Home sector is $\arg\max_s \overline{Z}$;
  $\mathrm{ratio}_Z$ and $\mathrm{ratio}_\beta$ are the in-home versus out-of-home concentration of
  $\overline{Z}$ and $|\overline{\beta}|$. The last column lists the three characteristics most
  correlated with the factor's long-short weight, with \textcolor{dkgreen}{$\uparrow$} for a positive
  and \textcolor{red}{$\downarrow$} for a negative correlation. Membership, loadings, and
  characteristics agree on a single theme per factor. Factor~0 is the equal-weight market, so its home
  sector and top characteristics are not meaningful and are left blank.}
\label{tab:factor_id}
\end{table}

\subsubsection{What drives each factor}

\begin{figure}[htbp]
\centering
 \begin{subfigure}[t]{0.63\textwidth}
   \centering
   \includegraphics[width=\textwidth]{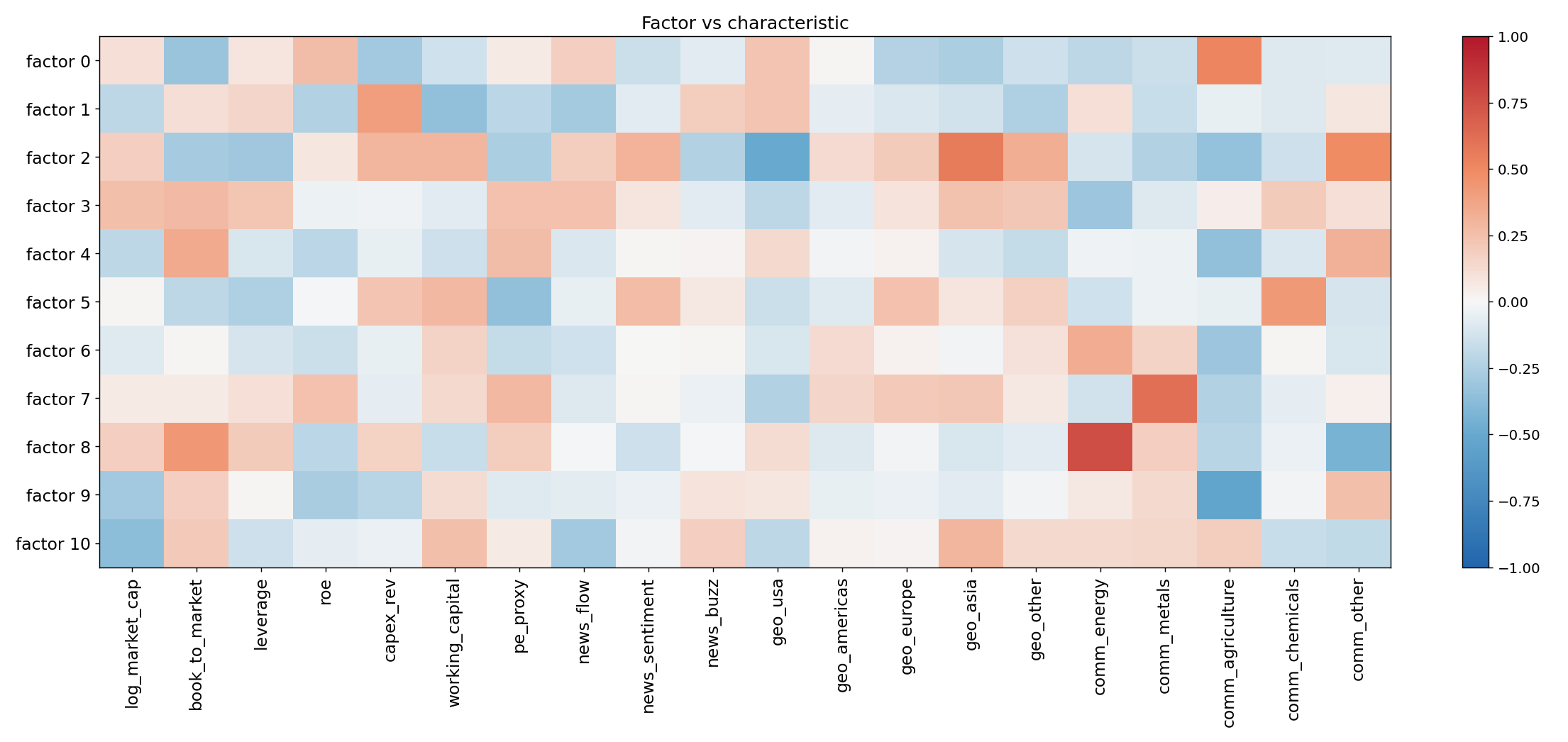}
   \caption{Factor--characteristic correlation.}
   \label{fig:factor_char_corr}
 \end{subfigure}
 \hfill
 \begin{subfigure}[t]{0.35\textwidth}
   \centering
   \raisebox{4mm}{\includegraphics[width=\textwidth]{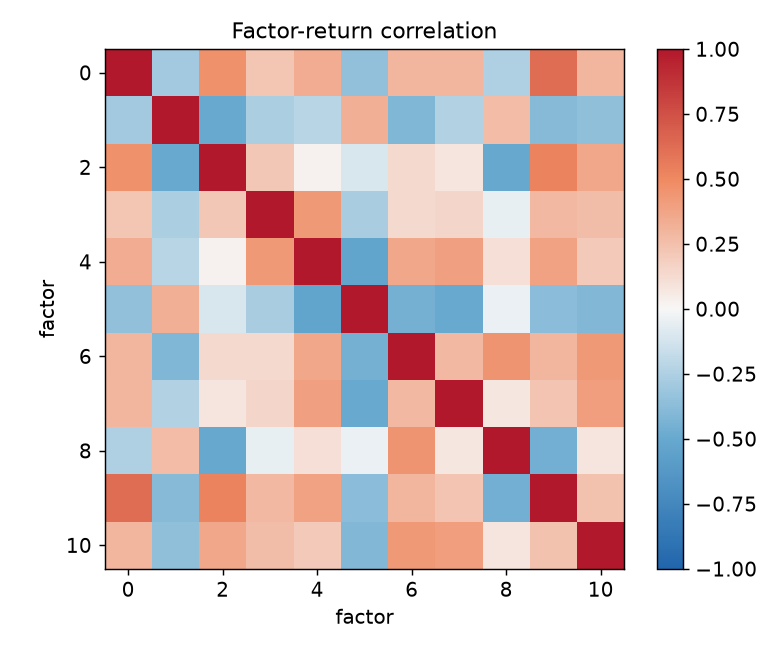}}
   \caption{Factor-return correlation.}
   \label{fig:factor_corr}
 \end{subfigure}
\caption{(a) Correlation of each factor with the input characteristics; factors align with
  recognisable characteristic groups rather than arbitrary directions. (b) Factor-return
  correlation matrix; low off-diagonal terms show the eleven factors span distinct directions rather
  than restating one another.}
\label{fig:factor_structure}
\end{figure}

\begin{figure}[htbp]
\centering
\includegraphics[width=0.75\textwidth]{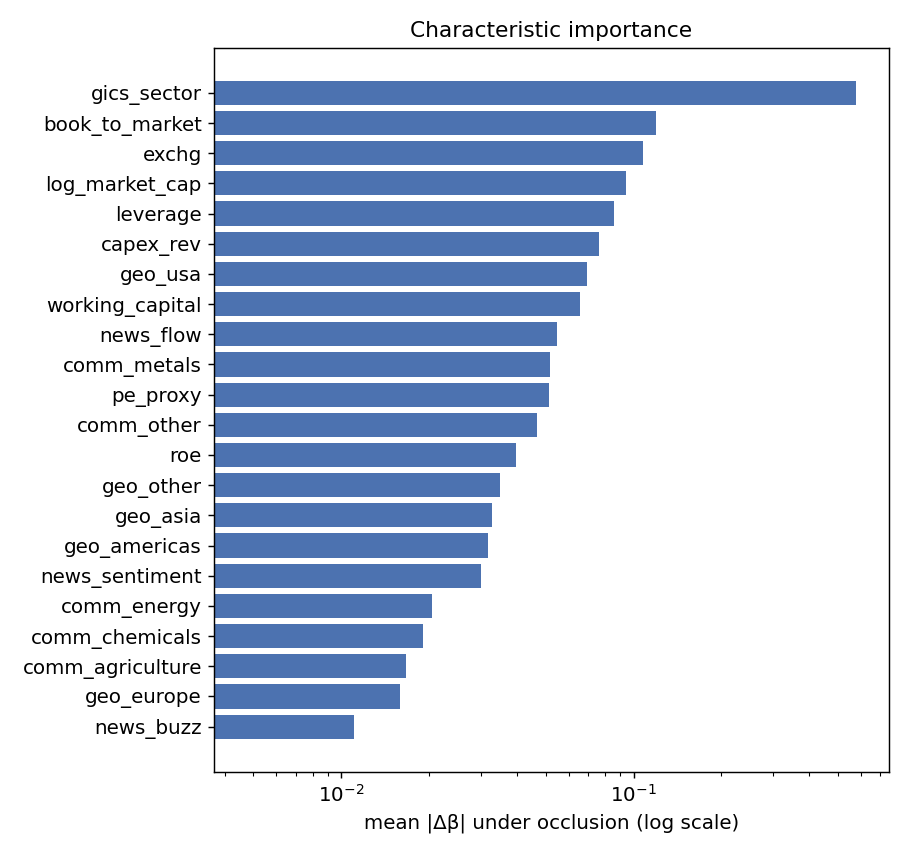}
\caption{Occlusion-based characteristic importance. Each bar is the mean change in the
  membership and loading heads when that characteristic is replaced by its cross-sectional neutral
  value, aggregated over all factors, so it measures how much a characteristic shapes the factor
  structure.}
\label{fig:feat_importance}
\end{figure}

The identities are not read off by hand. Figure~\ref{fig:factor_char_corr} correlates each factor with
the input characteristics, and each factor lines up with a narrow, recognisable group rather than a
diffuse blend. The occlusion importances \citep{zeiler2014visualizing} in
Figure~\ref{fig:feat_importance} tell the same story from the input side, measuring how far
neutralising each characteristic moves the membership and loading heads, they show that the factor
structure as a whole rests on a small, interpretable subset of the characteristics. Finally,
Figure~\ref{fig:factor_corr} shows the factor-return correlation matrix, whose off-diagonal terms are
low, so the eleven factors span distinct directions rather than restating one another.

\subsubsection{Behaviour across regimes}

\begin{figure}[htbp]
\centering
\includegraphics[width=\textwidth]{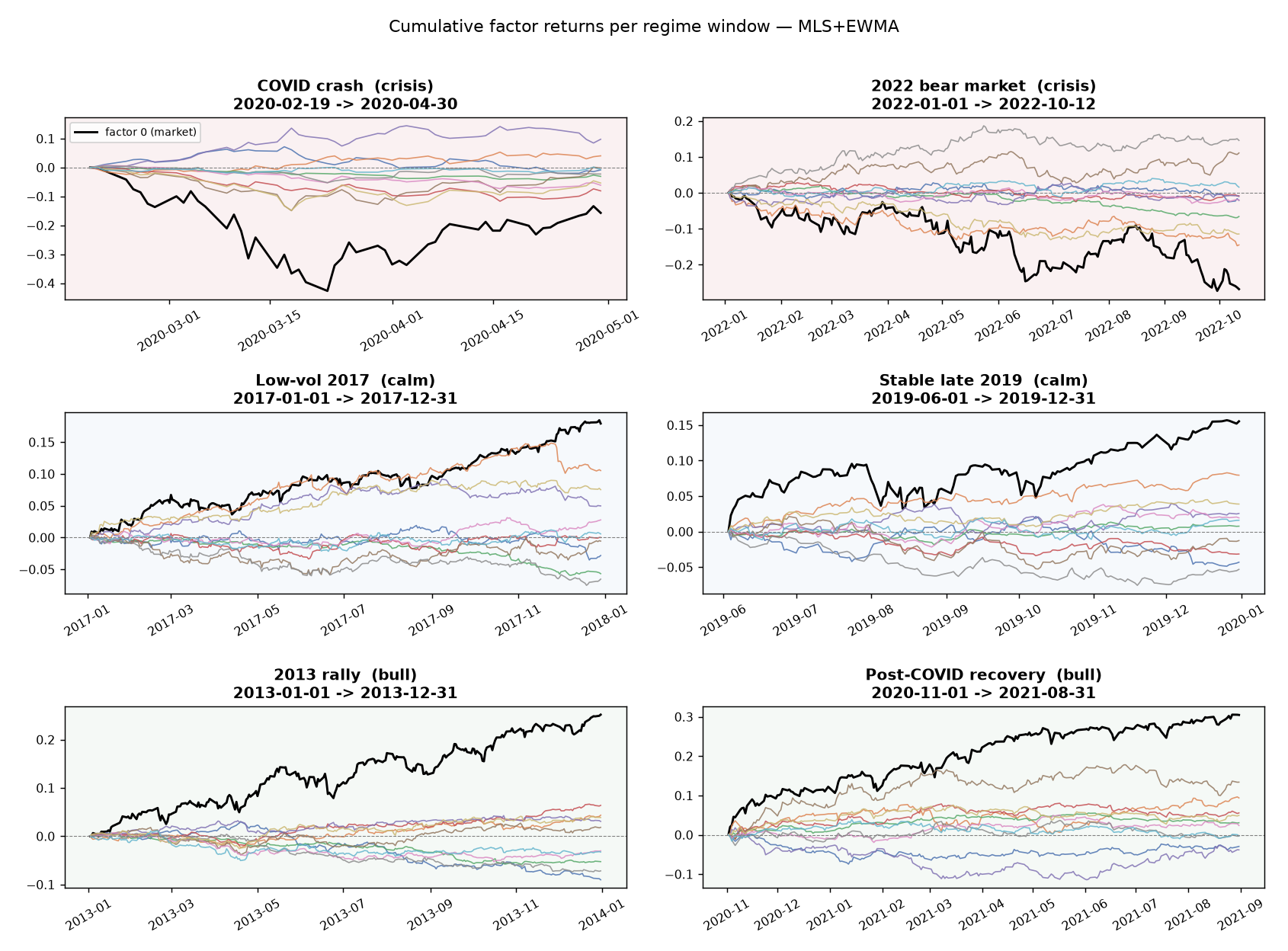}
\caption{Cumulative factor returns inside fixed historical regime windows. Crisis (COVID crash 2020,
  2022 bear), calm (2017, late 2019), and bull (2013, post-COVID 2020--21). The market factor (black)
  is the dominant, regime-dependent line; the long-short factors decouple defensively in the crises.}
\label{fig:regime_windows}
\end{figure}

A factor structure that is interpretable at one date should also stay coherent as the market moves.
Figure~\ref{fig:regime_windows} tracks the cumulative factor returns inside six fixed historical
windows, two crises (the COVID crash and the 2022 bear market), two calm periods, and two bull runs. The
market factor (black) is the dominant, strongly regime-dependent line, plunging in the crises and
carrying the bull runs, while the long-short factors decouple from it and reallocate risk across
regimes rather than simply moving together. The structure holds its shape through all three regime
types.

Which factors carry each regime is interpretable through Table~\ref{tab:factor_id}.
We present in Table~\ref{tab:regime_perf} the realised return of a few factors over the crisis and bull
windows. The market (factor~0) makes the directional swing, $-38\%$ in the crises and $+79\%$ in the
bull runs. Two defensive long-short factors stay positive when the market falls, utilities (factor~8)
and health care (factor~5), while the cyclical ones do the opposite, consumer discretionary
(factor~9), technology (factor~2) and financial value (factor~4) all losing in the crises and
recovering afterwards. Each factor moves in line with the sector it was assigned in
Table~\ref{tab:factor_id}, though the model was never told which sector to track.

\begin{table}[htbp]
\centering
\footnotesize
\setlength{\tabcolsep}{7pt}
\renewcommand{\arraystretch}{1.2}
\begin{tabular}{@{}c l cc l@{}}
\toprule
Factor & Home sector & Crisis & Bull & Role \\
\midrule
0 & Market (all)           & $-38\%$ & $+79\%$ & directional \\
\midrule
8 & Utilities              & $+13\%$ & $-7\%$  & defensive \\
5 & Health Care            & $+8\%$  & $-2\%$  & defensive \\
\midrule
9 & Consumer Discretionary & $-16\%$ & $+9\%$  & cyclical \\
2 & Information Technology  & $-10\%$ & $+14\%$ & cyclical \\
4 & Financials             & $-8\%$  & $+13\%$ & cyclical \\
\bottomrule
\end{tabular}
\caption{Realised factor return over the crisis (COVID crash and 2022 bear, 247 days) and bull
  (2013 and post-COVID recovery, 461 days) windows, for the market factor and the factors named in the
  text. Defensive factors gain when the market falls; cyclical ones track the market. Single seed;
  read for direction, not magnitude.}
\label{tab:regime_perf}
\end{table}

\subsubsection{Factor structure against benchmarks}
\label{sec:factor_bench}

To make interpretability measurable rather than asserted, we benchmark the factor structure of CD-DFM
against the competing models, and read where it stands. The full
suite is PCA, Sparse PCA \citep{zou2006}, IPCA
\citep{kelly2019}, and GKX \citep{gu2019}. The numbers are presented alongside the covariance results in
Table~\ref{tab:benchmarks} (Section~\ref{sec:covariance}).

\paragraph{Metrics.} Three quantities describe the factor structure and its leftover. Write
$e_{i,t} = r_{i,t} - \beta_{i,t}^\top f_t$ for the reconstruction residual.

\begin{itemize}[leftmargin=1.4em, topsep=2pt, itemsep=2pt, parsep=0pt]
  \item \emph{Reconstruction $R^2$}, the fraction of out-of-sample return variance the factors explain,
    $1 - \sum_{i,t} e_{i,t}^2 / \sum_{i,t} r_{i,t}^2$. Higher means the factors capture more of the
    systematic co-movement.

  \item \emph{Effective rank}, the participation ratio of the loading covariance eigenvalues,
    $(\sum_j \lambda_j)^2 / \sum_j \lambda_j^2$, in $[1, K]$. It counts how many factor directions are
    genuinely active. Read together with the next metric, a low value with low redundancy means a
    compact set of distinct factors, not collapse.

  \item \emph{Loading off-diagonal}, the mean absolute correlation between loading columns. A low value
    means the factors load on different stocks, so they are distinct rather than restating one another.
\end{itemize}

For the residual $e$, two metrics check what the factors leave behind.

\begin{itemize}[leftmargin=1.4em, topsep=2pt, itemsep=2pt, parsep=0pt]
  \item \emph{Residual cross-correlation}, the mean absolute off-diagonal correlation of the residuals
    across stocks. A low value means the factor structure has absorbed the systematic co-movement, and
    what remains is close to idiosyncratic.

  \item \emph{Residual PC1 share}, the variance share of the first principal component of the residuals.
    A low value means no single common factor was left unmodelled.
\end{itemize}

We also report the Ljung--Box $p$-value \citep{ljung1978} at ten lags on the residual mean series
(higher means the residual is closer to white noise, with less temporal structure left over).

\paragraph{Results.} On the factor axis, CD-DFM is the most parsimonious and the least redundant of the
five. It explains the most out-of-sample variance (reconstruction $R^2 = 0.526$ against $0.487$ to
$0.514$ for the rest) while using the fewest effective directions (effective rank $2.63$, near PCA's
$2.56$ and far below SPCA's $9.19$ or GKX's $6.13$) and the least collinear loadings (off-diagonal
$0.093$, roughly half of PCA's $0.128$ and a third of GKX's $0.234$). On the residual axis it leaves the
least systematic structure behind, with the lowest cross-correlation ($0.068$) at a comparable
common-variance level and the whitest residual of the suite (Ljung--Box $0.725$, against $0.632$ for GKX
and zeros for the static PCA models). A structured latent thus translates into a cleaner, more
idiosyncratic residual, which is what a good factor model should produce.

\subsection{Covariance estimation}
\label{sec:covariance}

The third property is the one that matters most in practice. A representation is only useful for risk if
it predicts the forward covariance out of sample, not merely if it reconstructs returns in sample. We
therefore evaluate the model's covariance $\hat\Sigma(t)$ against the realised forward second moment
$\Sigma^*(t)$ over the test period, and compare it to the full suite of competing estimators.

\paragraph{Baselines.} We benchmark against six estimators, from pure shrinkage to
characteristic-driven factor models. Ledoit--Wolf (LW) \citep{ledoit2004} and non-linear shrinkage (NLS)
\citep{ledoit2020} shrink the sample covariance directly and have no factors or latent space. PCA and
Sparse PCA \citep{zou2006} extract statistical factors from the return window. IPCA \citep{kelly2019}
and GKX \citep{gu2019} are the two characteristic-driven competitors, linear and non-linear
respectively, and are the closest in spirit to CD-DFM. The factor and residual metrics are defined only
for the five models that have a factor structure, so LW and NLS are blank on those rows.

\paragraph{Calibration metrics.} We present the metrics that score the covariance
$\hat\Sigma$ against the realised forward second moment $\Sigma^* = \frac1\Delta \sum_{s=t+1}^{t+\Delta}
r_s r_s^\top$, averaged over the test timestep. Let $\lambda_1(\cdot) \geq \dots \geq \lambda_n(\cdot)$
denote eigenvalues in decreasing order.

\begin{itemize}[leftmargin=1.4em, topsep=2pt, itemsep=2pt, parsep=0pt]
  \item \emph{Stein loss}, the Gaussian quasi-likelihood distance
    \[
      \mathcal{L}_{\text{Stein}}(\hat\Sigma, \Sigma^*)
      = \operatorname{tr}\!\big(\Sigma^* \hat\Sigma^{-1}\big)
      - \log\det\!\big(\Sigma^* \hat\Sigma^{-1}\big) - n .
    \]
    It is the training objective read out of sample, is zero only at $\hat\Sigma = \Sigma^*$, and
    penalises underestimated small eigenvalues heavily. Lower is better.

  \item \emph{Bures distance}, the Fr\'echet distance between the two covariances as operators,
    \[
      d_B(\hat\Sigma, \Sigma^*)
      = \Big( \operatorname{tr}\hat\Sigma + \operatorname{tr}\Sigma^*
        - 2\operatorname{tr}\big[(\hat\Sigma^{1/2}\Sigma^*\hat\Sigma^{1/2})^{1/2}\big] \Big)^{1/2},
    \]
    a purely geometric distance that does not go through a likelihood. Lower is better.

  \item \emph{Top eigenvalue ratio}, $\lambda_1(\hat\Sigma) / \lambda_1(\Sigma^*)$, which checks whether
    the dominant market-risk direction is scaled correctly. The target is $1$, and a value above $1$
    means an overstated top factor.

  \item \emph{Top-$K$ eigenvalue error}, the mean relative error on the leading $K$ eigenvalues,
    \[
      \frac1K \sum_{j=1}^{K}
      \frac{\big|\lambda_j(\hat\Sigma) - \lambda_j(\Sigma^*)\big|}{\lambda_j(\Sigma^*)} ,
    \]
    which looks past the top direction and asks whether the rest of the systematic spectrum is right.
    Lower is better.

  \item \emph{GMV and ERC calibration}. These are the practitioner's acid test for a covariance
    estimate. A risk model is ultimately used to build a portfolio, and what matters is whether the risk
    it predicts for that portfolio matches the risk actually realised. We form portfolio weights
    $w(\hat\Sigma)$ from the estimate and measure the calibration ratio
    \[
      \rho = \frac{w^\top \Sigma^*\, w}{w^\top \hat\Sigma\, w},
    \]
    the realised portfolio variance $w^\top \Sigma^*\, w$ over the ratio predicted at construction time
    $w^\top \hat\Sigma\, w$. We use two standard constructions that need only the covariance, not
    expected returns, as inputs~\citep{richard2015smart}:
    \begin{itemize}
        \item The \emph{Global Minimum Variance} (gmv) portfolio
        solves $\min_w w^\top \hat\Sigma\, w$ subject to $\mathbf{1}^\top w = 1$, giving
        $w_{\text{gmv}} = \hat\Sigma^{-1}\mathbf{1} / (\mathbf{1}^\top \hat\Sigma^{-1}\mathbf{1})$, and is the
        most sensitive to errors in the small eigenvalues.
        \item The \emph{Equal Risk Contribution} (erc) portfolio instead
        sets weights so that every asset contributes the same share $w_i (\hat\Sigma w)_i$ to portfolio
        variance, a construction that spreads the test across the whole matrix rather than its inverse. A
        value $\rho = 1$ is perfectly calibrated, $\rho > 1$ means the estimate understated risk, and
        $\rho < 1$ that it overstated it.
    \end{itemize}
\end{itemize}

\paragraph{Results.} Table~\ref{tab:benchmarks} collects the full comparison. CD-DFM leads the
covariance block, with the best Stein loss, top eigenvalue ratio and top-$K$ eigenvalue error, the
highest reconstruction $R^2$, and the whitest residual of the suite. What makes this notable is the
information the model works from. The characteristic-driven models estimate risk from slow-moving
fundamentals (Section~\ref{sec:data}), and with a mean lag-1 autocorrelation of about $0.68$ these
characteristics carry only a few hundred effectively independent states, several times fewer than the
roughly $700$ daily return observations available to PCA, Sparse PCA, LW and NLS.

From this much thinner information CD-DFM stays competitive with the return-based estimators. On GMV
calibration only PCA and Sparse PCA edge it ($0.97$ and $1.02$ against $0.93$), working from that larger
pool of returns, and even they fall behind on the equal-risk-contribution test where CD-DFM is the most
calibrated of the whole suite. NLS looks large at $1.42$ and $1.54$, but a ratio far from one is a
miscalibration that understates portfolio variance, not a strength.

The clearest gap is against our true competitors. IPCA and GKX are built from the same slow
characteristics as CD-DFM, and against them the advantage is decisive, for instance $0.93$ against their
$0.77$ on GMV calibration, on top of a more parsimonious and less redundant factor set. Taken together,
CD-DFM is the only model in the table at once predictive of covariance, parsimonious in its factors, and
interpretable against sectors, and it reaches this from characteristics alone.

\begin{table}[htbp]
\centering
\small
\setlength{\tabcolsep}{4pt}
\begin{tabular}{l C{1.25cm} C{1.25cm} C{1.25cm} C{1.25cm} C{1.25cm} C{1.25cm} C{1.25cm}}
\toprule
\textbf{Metric} & \textbf{CD-DFM} & \textbf{LW} & \textbf{NLS} & \textbf{PCA} & \textbf{SPCA} & \textbf{IPCA} & \textbf{GKX} \\
\midrule
\multicolumn{8}{l}{\textit{Calibration / covariance}} \\
Stein $\downarrow$        & \textbf{302} & 521 & 322 & 304 & \underline{303} & 309 & 309 \\
Bures $\downarrow$        & \underline{0.0262} & 0.0279 & 0.1197 & 0.0265 & \textbf{0.0261} & 0.0288 & 0.0277 \\
Top eig.\ ratio $\to 1$   & \textbf{1.64} & 1.91 & 5.61 & 1.98 & 1.96 & 1.84 & \underline{1.78} \\
Top-$K$ eig.\ err.\ $\downarrow$ & \textbf{0.26} & 0.38 & 0.74 & 0.49 & 0.38 & \underline{0.29} & 0.35 \\
$\rho_\text{gmv} \to 1$   & 0.93 & 0.92 & 1.42 & \underline{0.97} & \textbf{1.02} & 0.77 & 0.77 \\
$\rho_\text{erc} \to 1$   & \textbf{0.82} & \underline{0.73} & 1.54 & 0.71 & 0.73 & 0.71 & 0.71 \\
\midrule
\multicolumn{8}{l}{\textit{Fit / factors}} \\
Reconstruction $R^2$ $\uparrow$ & \textbf{0.526} & $\text{---}$ & $\text{---}$ & \underline{0.514} & 0.512 & 0.499 & 0.487 \\
Effective rank            & 2.63 & $\text{---}$ & $\text{---}$ & 2.56 & 9.19 & 4.16 & 6.13 \\
Loading off-diag.\ $\downarrow$ & \textbf{0.09} & $\text{---}$ & $\text{---}$ & \underline{0.13} & 0.21 & 0.16 & 0.23 \\
\midrule
\multicolumn{8}{l}{\textit{Residual quality}} \\
Cross-correlation $\downarrow$ & \textbf{0.068} & $\text{---}$ & $\text{---}$ & \underline{0.072} & 0.074 & 0.076 & 0.074 \\
PC1 share $\downarrow$    & 0.069 & $\text{---}$ & $\text{---}$ & \textbf{0.065} & \underline{0.067} & 0.085 & 0.074 \\
Ljung--Box $\uparrow$     & \textbf{0.73} & $\text{---}$ & $\text{---}$ & 0.00 & 0.00 & 0.57 & \underline{0.63} \\
\bottomrule
\end{tabular}
\caption{Full cross-benchmark comparison over the out-of-sample period (2022--2024). CD-DFM and GKX
  figures are averaged over 5 seeds, the others are single fits. Best per row in \textbf{bold}, second
  best \underline{underlined}; ``---'' marks metrics not defined for the factorless shrinkage estimators.
  The CD-DFM-vs-GKX latent comparison is reported separately in Table~\ref{tab:latent_gkx}.}
\label{tab:benchmarks}
\end{table}

\begin{figure}[htbp]
\centering
\begin{subfigure}{0.49\textwidth}
  \centering
  \includegraphics[width=\textwidth]{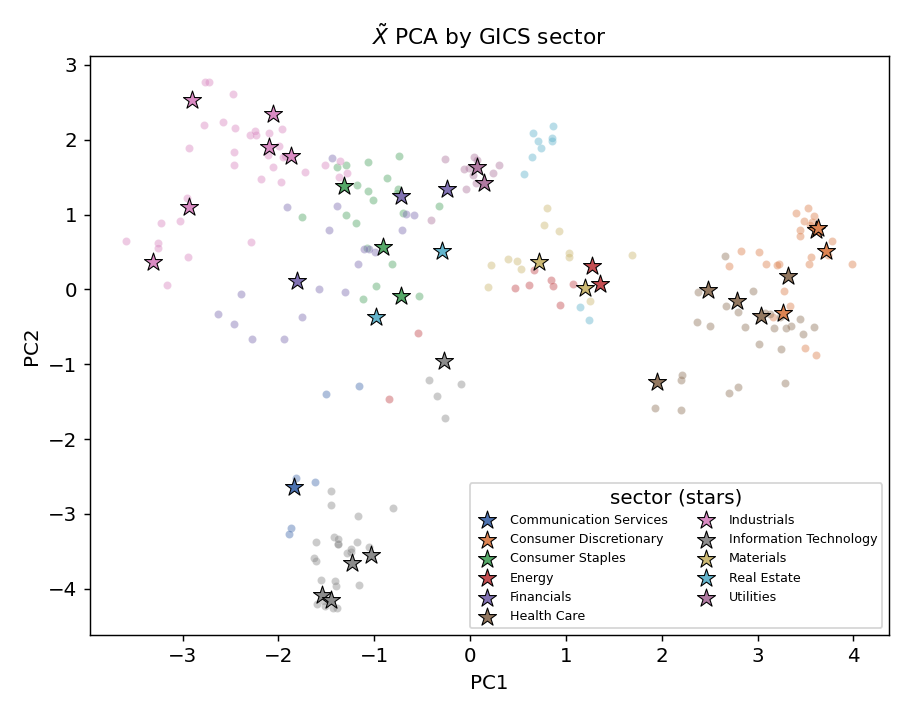}
  \caption{Held-out stocks in the latent space}
\end{subfigure}
\hfill
\begin{subfigure}{0.484\textwidth}
  \centering
  \includegraphics[width=\textwidth]{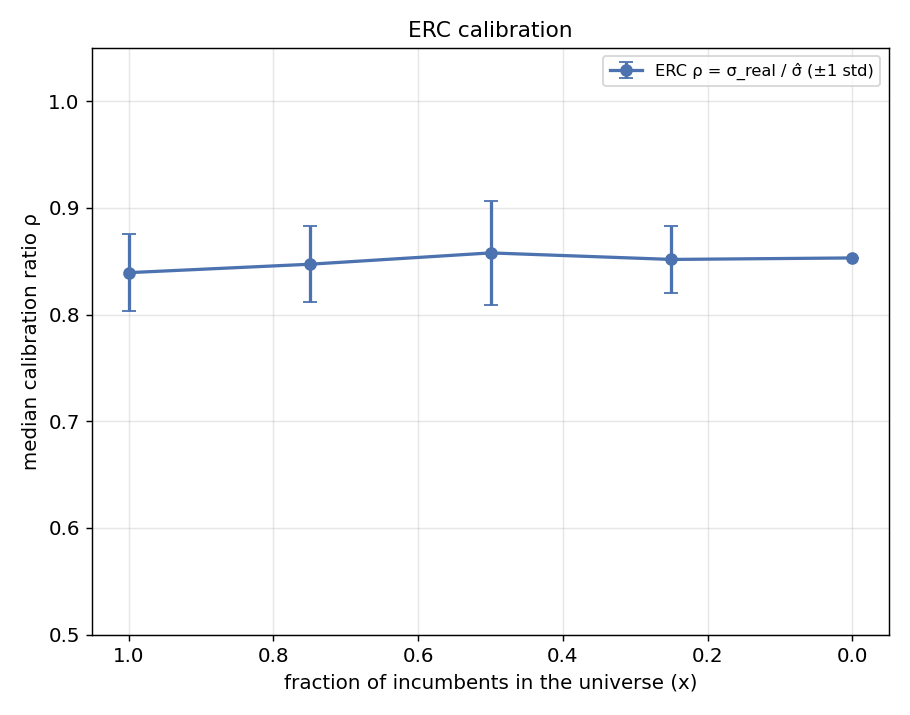}
  \caption{ERC calibration vs held-out fraction}
\end{subfigure}
\caption{Onboarding of unseen assets. Left, the $35$ held-out stocks are embedded from characteristics
  alone and land in their correct GICS sectors. Right, the equal-risk-contribution calibration $\rho$
  stays flat as the held-out fraction grows from $0$ to $100\%$, so calibration does not degrade for
  never-seen stocks.}
\label{fig:onboarding}
\end{figure}

\paragraph{Out of distribution generalisation: onboarding unseen assets.} Our characteristic-driven model can risk-model a stock it never trained
on, since the encoder maps characteristics to a latent position without needing past returns. We test
this directly by retraining on a reduced universe of $204$ incumbents and hold out $35$ stocks entirely,
then embed the held-out names at inference time and build the covariance with them included. Two things
should hold if onboarding works. The held-out stocks should land in the right region of the latent
space, and adding them should not degrade calibration. Both do.
Figure~\ref{fig:onboarding} shows the held-out names taking their correct sectoral position, and the
equal-risk-contribution calibration $\rho$ staying essentially flat as the held-out fraction rises from
$0$ to $100\%$ (from $0.853$ to $0.839$). The model onboards never-seen stocks with no parameter update
and no loss of calibration, which is the practical payoff of grounding the representation in
characteristics rather than in return history.

\section{Conclusion}

We introduced CD-DFM, a dynamic factor model driven solely by observable firm
characteristics that maps them to a latent representation of the equity cross-section, and from it to
interpretable factors and a forward covariance estimate. The model was designed around three
properties: (P1) a characteristic-driven latent space, (P2) interpretable factors, and (P3) an
accurate out-of-sample covariance. The empirical experiments over the 2022--2024 out-of-sample
period show it delivers all three. The latent geometry organises
the cross-section along economic lines and separates firms whose fundamentals differ even within a
sector. The factors carry recognisable economic themes and stay coherent across market regimes. The
covariance is the best calibrated of the benchmark suite taken jointly, and its residual is the whitest.

The central message is not that CD-DFM wins every metric in isolation, but that none of the benchmarks
combines all three properties as it does. What makes this notable is the input it works from: slow-moving
fundamentals rather than reactive return-based signals, and far fewer effectively distinct observations
than the return-based estimators. To our knowledge, it is the first model to recover forward risk from
slow-moving characteristics alone, which we believe is of direct practical value.
A further practical payoff of grounding the representation
in characteristics is its ability to generalise across the asset universe, that is, the
\emph{onboarding} of new assets, since a
never-seen stock is risk-modelled at inference time from its characteristics alone, with no parameter update
and no loss of calibration.

\paragraph{Limitations and future work.} The study is on a fixed universe of $239$ S\&P 500 stocks over
a single out-of-sample period, and a broader test across universes, longer horizons, and other asset
classes would strengthen the evidence. A more fundamental constraint is that a model of this kind is
data-hungry by design. Learning a non-linear characteristic-to-loading map end to end needs a long,
broad panel to be estimated reliably, and such rich, clean characteristic data is not always available,
which limits where the approach can be applied directly.

One direction we find promising is to use the characteristic-to-covariance map as a stress-testing tool.
Because the covariance is a function of the input characteristics, a user can write a scenario directly
in characteristic space, for instance a crisis expressed as a rise in news flow, a shift in geographic
or commodity dependence, and a move in valuation, and read off the covariance the model implies for that
scenario. This suggests using the model as a what-if tool for risk, where hypothetical states of the
world are described through the same interpretable inputs the model was trained on, rather than through
opaque shocks to the covariance itself. A systematic study of such characteristic-space stress tests,
and of how the implied risk compares to historical analogues, is left for future work.

\paragraph{Acknowledgments.}
The authors thank Paul Glasserman for early discussions on the use of alternative data for risk management.

This work was partially supported by the \emph{AI for Markets and Quantitative Investment} support fund of the \emph{Fondation de l'École Polytechnique}. The authors gratefully acknowledge the major donors of this fund: BNP Paribas, QRT, Squarepoint, and S\&P Global. This work was also carried out within the framework of a CIFRE PhD fellowship in collaboration with BNP Paribas Global Markets, Data \& AI Lab, and the Centre de Mathématiques Appliquées (CMAP), École polytechnique.

\newpage
\appendix

\section{Fast covariance inversion}
\label{app:woodbury}

The training loss needs $\hat\Sigma(t)^{-1}$ and $\log\det\hat\Sigma(t)$ at every timestep. Formed
directly these are $n \times n$ operations, cubic in the number of stocks. The low-rank-plus-diagonal
structure of \eqref{eq:lowrank} lets us compute both through a $K \times K$ system instead, with $K \ll
n$. Write $\hat\Sigma = \Psi + B B^\top$ with $\Psi = \operatorname{diag}(\psi)$ the idiosyncratic
variances and $B = \sqrt{\gamma}\,\beta\, \Sigma_f^{1/2} \in \mathbb{R}^{n \times K}$ the factor part.
The Woodbury identity gives the inverse,
\[
  \hat\Sigma^{-1}
  = \Psi^{-1} - \Psi^{-1} B \big(I_K + B^\top \Psi^{-1} B\big)^{-1} B^\top \Psi^{-1},
\]
where the only matrix inverted is the $K \times K$ capacitance $I_K + B^\top \Psi^{-1} B$, since
$\Psi$ is diagonal. The matrix-determinant lemma gives the log-determinant on the same $K \times K$
object,
\[
  \log\det \hat\Sigma
  = \log\det\!\big(I_K + B^\top \Psi^{-1} B\big) + \sum_{i=1}^{n} \log \psi_i .
\]
Both cost $O(nK^2 + K^3)$, linear in $n$. The quadratic form $r^\top \hat\Sigma^{-1} r$ in the Stein
loss is then evaluated without ever materialising the dense $\hat\Sigma^{-1}$, and $\Psi$ being strictly
positive keeps the capacitance matrix positive definite, so $\hat\Sigma$ is invertible by construction
even when the factor block is low rank.

\section{Architecture and hyperparameters}
\label{app:hyperparams}

The encoder follows the FT-Transformer style of tabular encoding. Each numeric
characteristic $j$ (continuous or compositional) is mapped to a token by a per-feature affine map
$x_j \mapsto x_j w_j + b_j \in \mathbb{R}^d$, and each categorical (GICS sector, exchange) is mapped to a
$d$-dimensional embedding through its own table. The tokens are concatenated and passed through a
two-layer MLP to the latent embedding $\tilde X \in \mathbb{R}^{n \times D}$. The membership head applies
a softmax to give $Z \in \Delta^{K-1}$, and the loading head is a separate linear map
to $\beta \in \mathbb{R}^{n \times K}$. Table~\ref{tab:hyperparams} lists the settings of the retained
configuration.

\begin{table}[htbp]
\centering
\small
\begin{tabular}{l l l}
\toprule
\textbf{Group} & \textbf{Parameter} & \textbf{Value} \\
\midrule
\multirow{4}{*}{Model}
  & latent dimension $D$ & $64$ \\
  & factors $K$ & $11$ (Marcenko--Pastur on train) \\
  & token dimension $d$ & $8$ \\
  & MLP hidden width & $128$ \\
\midrule
\multirow{4}{*}{Factors / covariance}
  & factor construction & market $+$ $K{-}1$ long-short, $\ell_1$-normalised \\
  & $\Sigma_f$ EWMA half-life & $90$ days \\
  & loss weight $\alpha$ & $0.7$ \\
\midrule
\multirow{3}{*}{Windows}
  & lookback $H$ & $252$ days \\
  & forward horizon $\Delta$ & $252$ days \\
  & anchor stride & $5$ days \\
\midrule
\multirow{5}{*}{Optimisation}
  & optimiser & Adam, learning rate $10^{-3}$ \\
  & batch size (timesteps) & $32$ \\
  & max epochs / patience & $100$ / $10$ \\
  & early-stopping criterion & validation Stein loss \\
\bottomrule
\end{tabular}
\caption{Hyperparameters of the retained configuration.}
\label{tab:hyperparams}
\end{table}

\section{Training practices}
\label{app:training}

A few choices made training stable and are worth stating for reproducibility. Before training we
calibrate the per-term scales of the objective on a small set of timesteps, so the reconstruction and
covariance terms enter $\mathcal{L}$ on a comparable magnitude and $\alpha$ acts as a genuine
weight rather than being dominated by a scale mismatch. The train, validation and test splits are
chronological with a $\Delta$-day embargo between them, so no training forward window reaches into the
validation or test period. The number of factors $K$ is fixed once, before training, by a
Marcenko--Pastur threshold on the training return block, rather than tuned on the covariance metrics.
Concretely, for the $T \times N$ block of training returns (with $T > N$) and aspect ratio $q = N/T$,
the correlation eigenvalues of pure noise fall inside the Marcenko--Pastur support \citep{marcenko1967}
with upper edge $\lambda_+ = (1 + \sqrt{q})^2$. We set $K$ to the number of correlation eigenvalues that
exceed $\lambda_+$, since those carry cross-sectional structure rather than noise, which gives $K = 11$
here.

The factor covariance $\Sigma_f$ is estimated with the exponentially weighted moving average
\citep{riskmetrics1996} of \eqref{eq:sigmaf}, so it tracks the current volatility regime; here we give the
exact weighting. Ordering the lookback days by age $a = 0, \dots, H-1$ from most to least recent, the
weights decay geometrically,
\[
  w_a \propto \lambda^{a}, \qquad \lambda = 2^{-1/h_{1/2}}, \qquad
  \Sigma_f = \sum_{a=0}^{H-1} w_a\, f_a f_a^\top,
\]
with the weights normalised to sum to one and $h_{1/2}$ the half-life, the age at which a day's weight
halves. A short half-life reacts fast but is noisier, a long one is smoother but slower, and
$h_{1/2} = 90$ days is the value retained (Appendix~\ref{app:ablations}). The recency weighting is
applied to $\Sigma_f$ only, while the idiosyncratic variances $\Psi$ keep the flat lookback average, so
the factor block follows the regime without making the residual variances noisy.

\section{Ablations}
\label{app:ablations}

\paragraph{Reconstruction versus covariance weight $\alpha$.} The objective mixes a reconstruction term
and a covariance term with weight $\alpha$ (\eqref{eq:ltot}). Table~\ref{tab:ablation_alpha} compares the
two extremes against the retained $\alpha = 0.7$. Pure covariance ($\alpha = 1$) drops the reconstruction
$R^2$ sharply and overstates the top eigenvalue, since nothing anchors the factors to the returns they
should reconstruct. Pure reconstruction ($\alpha = 0$) keeps a good $R^2$ but leaves more redundant
factors and a weaker ERC calibration. The mixed objective is the balance that keeps both sides healthy.

\begin{table}[htbp]
\centering
\small
\begin{tabular}{l C{2.4cm} C{2.4cm} C{2.4cm}}
\toprule
\textbf{Metric} & $\alpha = 0$ (recon) & $\alpha = 0.7$ & $\alpha = 1$ (cov) \\
\midrule
Reconstruction $R^2$ $\uparrow$ & 0.52 & \textbf{0.53} & 0.41 \\
Top eig.\ ratio $\to 1$         & 1.69 & \textbf{1.72} & 1.95 \\
Loading off-diag.\ $\downarrow$ & 0.15 & \textbf{0.09} & 0.14 \\
$\rho_\text{erc} \to 1$         & 0.79 & \textbf{0.80} & 0.76 \\
\bottomrule
\end{tabular}
\caption{Effect of the reconstruction/covariance weight $\alpha$ (retained configuration, single seed).
  These are seed-0 fits, so shared metrics differ slightly from the 5-seed means in
  Table~\ref{tab:benchmarks}.}
\label{tab:ablation_alpha}
\end{table}

\paragraph{Recency weighting of $\Sigma_f$.} The factor covariance $\Sigma_f$ is estimated with an EWMA
half-life over the lookback window. Table~\ref{tab:ablation_ewma} compares the retained $90$-day
half-life to shorter ones and to a flat average. A short half-life reacts faster but calibrates slightly
worse, while the flat average clearly understates portfolio risk, with the lowest GMV and ERC
calibration and the most overstated top eigenvalue. The $90$-day half-life is the best joint
calibration.

\begin{table}[htbp]
\centering
\small
\begin{tabular}{l C{1.9cm} C{1.9cm} C{1.9cm} C{1.9cm}}
\toprule
\textbf{Metric} & 10 days & 15 days & \textbf{90 days} & flat \\
\midrule
$\rho_\text{gmv} \to 1$   & 0.92 & 0.90 & \textbf{0.93} & 0.87 \\
$\rho_\text{erc} \to 1$   & 0.80 & 0.79 & \textbf{0.80} & 0.72 \\
Top eig.\ ratio $\to 1$   & \textbf{1.66} & 1.65 & 1.72 & 1.85 \\
Reconstruction $R^2$ $\uparrow$ & 0.52 & 0.53 & 0.53 & 0.53 \\
\bottomrule
\end{tabular}
\caption{Effect of the $\Sigma_f$ EWMA half-life (retained configuration, single seed). ``flat'' is a
  uniform window average. These are seed-0 fits, so shared metrics differ slightly from the 5-seed means
  in Table~\ref{tab:benchmarks}.}
\label{tab:ablation_ewma}
\end{table}

\section{A volatility-reconstruction variant}
\label{app:vol}

The reconstruction term in \eqref{eq:lrec} fits the signed returns, $\lVert r_s - \beta_s f_s\rVert^2$.
A natural risk-oriented alternative is to fit the squared magnitude instead,
\begin{equation}\label{eq:lvol}
  \mathcal{L}_{\mathrm{vol}}(t) = \frac1{H}\sum_{s=t-H+1}^{t}
     \bigl\lVert r_s^{2} - (\beta_s f_s)^{2} \bigr\rVert^2 ,
\end{equation}
where the square acts componentwise. Matching $r_s^{2}$ to $(\beta_s f_s)^{2}$ makes the term
sign-blind, so it scores how well the systematic block explains the amplitude of returns rather than
their direction. The whole objective then targets second moments only, which is conceptually clean for a
risk model that is never asked to predict the sign of a return.

We report this variant for completeness. Table~\ref{tab:vol_variant} compares it to the signed CD-DFM,
each averaged over the same five seeds. The two are close on the covariance and calibration rows, and
the volatility variant is in fact a little stronger on the latent structure, with a higher sector and
loading silhouette. It gives up some ground on the signed-fit metrics, the reconstruction $R^2$ and the
residual whiteness, which is expected since it never fits the signed level. The more striking reading is
the reverse: the volatility variant reconstructs returns from their magnitudes alone, never seeing the
sign of $r$, yet stays within a few points of the signed model on reconstruction $R^2$ and matches it
across the covariance and calibration metrics. This says that most of the cross-sectional risk structure,
the loadings and the resulting covariance, lives in the scale of returns rather than their direction, and
can be recovered from volatilities alone. We keep the signed reconstruction as the default for its
sharper fit and whiter residual, and note the volatility variant as an elegant, purely second-moment
alternative that performs essentially on par.

\begin{table}[htbp]
\centering
\small
\begin{tabular}{l C{2.6cm} C{2.6cm}}
\toprule
\textbf{Metric} & \textbf{CD-DFM (signed)} & Volatility variant \\
\midrule
\multicolumn{3}{l}{\textit{Calibration / covariance}} \\
Stein $\downarrow$              & \textbf{302} & 310 \\
Top eig.\ ratio $\to 1$         & \textbf{1.64} & 1.70 \\
Top-$K$ eig.\ err.\ $\downarrow$ & \textbf{0.26} & 0.28 \\
$\rho_\text{gmv} \to 1$         & \textbf{0.93} & 0.92 \\
$\rho_\text{erc} \to 1$         & \textbf{0.82} & 0.80 \\
\midrule
\multicolumn{3}{l}{\textit{Fit / residual}} \\
Reconstruction $R^2$ $\uparrow$ & \textbf{0.53} & 0.48 \\
Loading off-diag.\ $\downarrow$ & \textbf{0.09} & 0.14 \\
Ljung--Box $\uparrow$           & \textbf{0.73} & 0.32 \\
\midrule
\multicolumn{3}{l}{\textit{Latent structure}} \\
Silhouette (sector) $\uparrow$  & 0.38 & \textbf{0.40} \\
Silhouette ($\beta$) $\uparrow$ & 0.36 & \textbf{0.40} \\
\bottomrule
\end{tabular}
\caption{Signed CD-DFM versus the volatility-reconstruction variant \eqref{eq:lvol}, both averaged over
  five seeds. Best per row in \textbf{bold}. The two are comparable, the signed default sharper on the
  fit and residual rows, the volatility variant slightly more structured in the latent space.}
\label{tab:vol_variant}
\end{table}

\bibliographystyle{plainnat}
\bibliography{references}

@article{ljung1978,
  author  = {Ljung, Greta M. and Box, George E. P.},
  title   = {On a Measure of Lack of Fit in Time Series Models},
  year    = {1978},
  journal = {Biometrika}
}

@article{spearman1904,
  author  = {Spearman, Charles},
  title   = {The Proof and Measurement of Association between Two Things},
  year    = {1904},
  journal = {The American Journal of Psychology}
}

@article{jaccard1912,
  author  = {Jaccard, Paul},
  title   = {The Distribution of the Flora in the Alpine Zone},
  year    = {1912},
  journal = {New Phytologist}
}

@inproceedings{yang2026,
  author  = {Yang, Jie and Hu, Yifan and Li, Yuante and Zhang, Kexin and Ding, Kaize and Yu, Philip S.},
  title   = {From Observations to States: Latent Time Series Forecasting},
  year    = {2026},
  booktitle = {International Conference on Machine Learning}
}

@article{gu2019,
  author    = {Gu, Shihao and Kelly, Bryan and Xiu, Dacheng},
  title     = {Autoencoder Asset Pricing Models},
  year      = {2019},
  journal   = {Chicago Booth Working Paper No.\ 19-24}
}

@article{marcenko1967,
  author  = {Mar\v{c}enko, Vladimir A. and Pastur, Leonid A.},
  title   = {Distribution of Eigenvalues for Some Sets of Random Matrices},
  year    = {1967},
  journal = {Mathematics of the USSR-Sbornik}
}

@techreport{riskmetrics1996,
  author      = {{J.P. Morgan}},
  title       = {RiskMetrics Technical Document},
  year        = {1996},
  institution = {J.P. Morgan/Reuters},
  edition     = {4th}
}

@article{ledoit2004,
  author  = {Ledoit, Olivier and Wolf, Michael},
  title   = {A Well-Conditioned Estimator for Large-Dimensional Covariance Matrices},
  year    = {2004},
  journal = {Journal of Multivariate Analysis}
}

@article{ledoit2020,
  author  = {Ledoit, Olivier and Wolf, Michael},
  title   = {Analytical Nonlinear Shrinkage of Large-Dimensional Covariance Matrices},
  year    = {2020},
  journal = {The Annals of Statistics}
}

@article{zou2006,
  author  = {Zou, Hui and Hastie, Trevor and Tibshirani, Robert},
  title   = {Sparse Principal Component Analysis},
  year    = {2006},
  journal = {Journal of Computational and Graphical Statistics}
}

@article{kelly2019,
  author  = {Kelly, Bryan T. and Pruitt, Seth and Su, Yinan},
  title   = {Characteristics Are Covariances: A Unified Model of Risk and Return},
  year    = {2019},
  journal = {Journal of Financial Economics}
}

@article{benzecri1977analyse,
  title={Analyse discriminante et analyse factorielle},
  author={Benz{\'e}cri, J-P},
  journal={Les cahiers de l'analyse des donn{\'e}es},
  volume={2},
  number={4},
  pages={369--406},
  year={1977}
}

@book{deboeck2013visual,
  title={Visual explorations in finance: with self-organizing maps},
  author={Deboeck, Guido and Kohonen, Teuvo},
  year={2013},
  publisher={Springer Science \& Business Media}
}

@article{bengio2013,
  author  = {Bengio, Yoshua and Courville, Aaron and Vincent, Pascal},
  title   = {Representation Learning: A Review and New Perspectives},
  year    = {2013},
  journal = {IEEE Transactions on Pattern Analysis and Machine Intelligence}
}

@article{welch2008comprehensive,
  title={A comprehensive look at the empirical performance of equity premium prediction},
  author={Welch, Ivo and Goyal, Amit},
  journal={The Review of Financial Studies},
  volume={21},
  number={4},
  pages={1455--1508},
  year={2008},
  publisher={Society for Financial Studies}
}

@article{harvey2016and,
  title={… and the cross-section of expected returns},
  author={Harvey, Campbell R and Liu, Yan and Zhu, Heqing},
  journal={The Review of financial studies},
  volume={29},
  number={1},
  pages={5--68},
  year={2016},
  publisher={Oxford University Press}
}

@inproceedings{gorishniy2021,
  author  = {Gorishniy, Yury and Rubachev, Ivan and Khrulkov, Valentin and Babenko, Artem},
  title   = {Revisiting Deep Learning Models for Tabular Data},
  year    = {2021},
  booktitle = {Advances in Neural Information Processing Systems}
}

@incollection{ross2013arbitrage,
  title={The arbitrage theory of capital asset pricing},
  author={Ross, Stephen A},
  booktitle={Handbook of the fundamentals of financial decision making: Part I},
  pages={11--30},
  year={2013},
  publisher={World Scientific}
}

@incollection{WynneBrownCeria2014,
  author    = {Wynne, Bill and Brown, Melissa and Ceria, Sebastian},
  title     = {Axioma Risk Models},
  booktitle = {Investment Risk and Uncertainty: Advanced Risk Awareness Techniques for the Intelligent Investor},
  editor    = {Greiner, Michael P.},
  publisher = {Wiley},
  year      = {2014}
}

@manual{MSCIBarraGEM3,
  title  = {Barra Global Equity Model (GEM3) Methodology},
  author = {{MSCI}},
  year   = {2013},
  note   = {Methodology document, MSCI Barra}
}

@article{cetingoz2026synthetic,
  title={Synthetic data for portfolios: A throw of the dice will never abolish chance},
  author={Cetingoz, Adil Rengim and Lehalle, Charles-Albert},
  journal={Quantitative Finance},
  pages={1--32},
  year={2026},
  publisher={Taylor \& Francis}
}

@inproceedings{james1961estimation,
  title={Estimation with quadratic loss},
  author={James, William and Stein, Charles},
  booktitle={Proceedings of the fourth Berkeley symposium on mathematical statistics and probability},
  volume={1},
  number={1961},
  pages={361--379},
  year={1961},
  organization={University of California Press}
}

@article{demiguel2009generalized,
  title={A generalized approach to portfolio optimization: Improving performance by constraining portfolio norms},
  author={DeMiguel, Victor and Garlappi, Lorenzo and Nogales, Francisco J and Uppal, Raman},
  journal={Management science},
  volume={55},
  number={5},
  pages={798--812},
  year={2009},
  publisher={INFORMS}
}

@inproceedings{zeiler2014visualizing,
  title={Visualizing and understanding convolutional networks},
  author={Zeiler, Matthew D and Fergus, Rob},
  booktitle={European conference on computer vision},
  pages={818--833},
  year={2014},
  organization={Springer}
}

@article{fefferman2016,
  author  = {Fefferman, Charles and Mitter, Sanjoy and Narayanan, Hariharan},
  title   = {Testing the Manifold Hypothesis},
  year    = {2016},
  journal = {Journal of the American Mathematical Society}
}

@incollection{richard2015smart,
  title={Smart beta: Managing diversification of minimum variance portfolios},
  author={Richard, Jean-Charles and Roncalli, Thierry},
  booktitle={Risk-Based and Factor Investing},
  pages={31--63},
  year={2015},
  publisher={Elsevier}
}

@article{fan2013,
  author  = {Fan, Jianqing and Liao, Yuan and Mincheva, Martina},
  title   = {Large Covariance Estimation by Thresholding Principal Orthogonal Complements},
  year    = {2013},
  journal = {Journal of the Royal Statistical Society: Series B}
}

\end{document}